\documentclass[apj]{emulateapj}

\usepackage{apjfonts}
\usepackage{amssymb, amsmath}
\usepackage{natbib}
\usepackage{ifthen}
\usepackage{appendix}
\newcommand\degree{\degr}
\newcommand\degrees\degree
\newcommand\vs{{\em vs.}\ }

\DeclareSymbolFont{UPM}{U}{eur}{m}{n}
\DeclareMathSymbol{\umu}{0}{UPM}{"16}
\let\oldumu=\umu
\renewcommand\umu{\ifmmode\oldumu\else\math{\oldumu}\fi}
\newcommand\micro{\umu}
\renewcommand\micron{\micro m}
\newcommand\microns \micron

\renewcommand\arcsec[0]{$^{\prime\prime}$}

\let\oldsim=\sim
\renewcommand\sim{\ifmmode\oldsim\else\math{\oldsim}\fi}
\let\oldpm=\pm
\renewcommand\pm{\ifmmode\oldpm\else\math{\oldpm}\fi}
\newcommand\by{\ifmmode\times\else\math{\times}\fi}

\newbox{\wdbox}
\renewcommand\c{\setbox\wdbox=\hbox{,}\hspace{\wd\wdbox}}
\renewcommand\i{\setbox\wdbox=\hbox{i}\hspace{\wd\wdbox}}

\newcount\timect
\newcount\hourct
\newcount\minct
\newcommand\now{\timect=\time \divide\timect by 60
         \hourct=\timect \multiply\hourct by 60
         \minct=\time \advance\minct by -\hourct
         \number\timect:\ifnum \minct < 10 0\fi\number\minct}

\newcommand\mctc{\multicolumn{2}{c}}

\catcode`@=11

\newcommand\comment[1]{}
\newcommand\commenton{\catcode`\%=14}
\newcommand\commentoff{\catcode`\%=12}

\renewcommand\math[1]{$#1$}
\newcommand\mathshifton{\catcode`\$=3}
\newcommand\mathshiftoff{\catcode`\$=12}

\comment{the backslash is necessary}

\comment{alignment tab}

\let\atab=&
\newcommand\atabon{\catcode`\&=4}
\newcommand\ataboff{\catcode`\&=12}

\let\oldmsp=\sp
\let\oldmsb=\sb
\def\sp#1{\ifmmode
           \oldmsp{#1}%
         \else\strut\raise.85ex\hbox{\scriptsize #1}\fi}
\def\sb#1{\ifmmode
           \oldmsb{#1}%
         \else\strut\raise-.54ex\hbox{\scriptsize #1}\fi}
\newbox\@sp
\newbox\@sb
\def\sbp#1#2{\ifmmode%
           \oldmsb{#1}\oldmsp{#2}%
         \else
           \setbox\@sb=\hbox{\sb{#1}}%
           \setbox\@sp=\hbox{\sp{#2}}%
           \rlap{\copy\@sb}\copy\@sp
           \ifdim \wd\@sb >\wd\@sp
             \hskip -\wd\@sp \hskip \wd\@sb
           \fi
        \fi}
\def\msp#1{\ifmmode
           \oldmsp{#1}
         \else \math{\oldmsp{#1}}\fi}
\def\msb#1{\ifmmode
           \oldmsb{#1}
         \else \math{\oldmsb{#1}}\fi}
\def\supon{\catcode`\^=7}
\def\supoff{\catcode`\^=12}
\def\subon{\catcode`\_=8}
\def\suboff{\catcode`\_=12}
\def\supsubon{\supon \subon}
\def\supsuboff{\supoff \suboff}

\newcommand\actcharon{\catcode`\~=13}
\newcommand\actcharoff{\catcode`\~=12}

\newcommand\paramon{\catcode`\#=6}
\newcommand\paramoff{\catcode`\#=12}

\comment{And now to turn us totally on and off...}

\newcommand\reservedcharson{\commenton \mathshifton \atabon \supsubon \actcharon
	\paramon}

\newcommand\reservedcharsoff{\commentoff \mathshiftoff \ataboff
	\supsuboff \actcharoff \paramoff}

\catcode`@=12
\reservedcharsoff

\reservedcharson

\comment{ Must have ONLY ONE of these... trust these macros, they work

}

\newcommand{\squishlist}{
 \begin{list}{$\bullet$}
  { \setlength{\itemsep}{1pt}
     \setlength{\parsep}{0pt}
     \setlength{\topsep}{3pt}
     \setlength{\partopsep}{0pt}
     \setlength{\leftmargin}{2.0em}
     \setlength{\labelwidth}{1.5em}
     \setlength{\labelsep}{0.5em} } }

\newcommand{\squishend}{
  \end{list}  }

\reservedcharson

\actcharon

\reservedcharson
\actcharon

\submitted{Accepted by ApJ 2012-06-07}

\shorttitle{Two nearby sub-Earth-sized exoplanet candidates in the GJ 436 system}
\shortauthors{Stevenson {\em et al.}}

\begin{document}

\title{Two nearby sub-Earth-sized exoplanet candidates in the GJ 436 system}

\author{Kevin B.\ Stevenson, Joseph Harrington, and Nate B.\ Lust}
\affil{Planetary Sciences Group, Department of Physics, University of Central Florida\\
Orlando, FL 32816-2385, USA}

\author{Nikole K.\ Lewis}
\affil{Department of Planetary Sciences and Lunar and Planetary Laboratory, The University of Arizona, Tucson, AZ 85721, USA}

\author{Guillaume Montagnier}
\affil{European Organisation for Astronomical Research in the Southern Hemisphere (ESO), Casilla 19001, Santiago 19, Chile}

\author{Julianne I.\ Moses}
\affil{Space Science Institute, 4750 Walnut St, Suite 205, Boulder, CO 80301, USA}

\author{Channon Visscher}
\affil{Southwest Research Institute, 1050 Walnut St., Suite 300, Boulder CO, 80302, USA}

\author{Jasmina Blecic, Ryan A.\ Hardy, Patricio Cubillos, and Christopher J.\ Campo}
\affil{Planetary Sciences Group, Department of Physics, University of Central Florida\\
Orlando, FL 32816-2385, USA}

\email{kevin218@knights.ucf.edu}

\begin{abstract}

We report the detection of UCF-1.01, a strong exoplanet candidate with a radius 0.66 {\pm} 0.04 times that of Earth ($R\sb{\oplus}$).  This sub-Earth-sized planet transits the nearby M-dwarf star GJ 436 with a period of 1.365862 {\pm} 8$\times$10$^{-6}$ days.  We also report evidence of a 0.65 {\pm} 0.06 $R\sb{\oplus}$ exoplanet candidate (labeled UCF-1.02) orbiting the same star with an undetermined period.  Using the {\em Spitzer Space Telescope}, we measure the dimming of light as the planets pass in front of their parent star to assess their sizes and orbital parameters. 
If confirmed, UCF-1.01 and UCF-1.02 would be called GJ 436c and GJ 436d, respectively, and would be part of the first multiple-transiting-planet system outside of the Kepler field. 
Assuming Earth-like densities of 5.515 g/cm$^{3}$, we predict both candidates to have similar masses ($\sim$0.28 Earth-masses, $M\sb{\oplus}$, 2.6 Mars-masses) and surface gravities of $\sim$0.65 $g$ (where $g$ is the gravity on Earth).
UCF-1.01's equilibrium temperature ($T$\sb{eq}, where emitted and absorbed radiation balance for an equivalent blackbody) is 860 K, making the planet unlikely to harbor life as on Earth.  Its weak gravitational field and close proximity to its host star imply that UCF-1.01 is unlikely to have retained its original atmosphere; however, a transient atmosphere is possible if recent impacts or tidal heating were to supply volatiles to the surface.
We also present additional observations of GJ 436b during secondary eclipse.  The 3.6-{\micron} light curve shows indications of stellar activity, making a reliable secondary eclipse measurement impossible.  A second non-detection at 4.5 {\microns} supports our previous work in which we find a methane-deficient and carbon monoxide-rich dayside atmosphere.

\end{abstract}

\section{INTRODUCTION}
\label{intro}

The search for Earth-sized planets around main-sequence stars has progressed expeditiously in the last year.  Recent discoveries include two Earth-sized planets (0.868 and 1.03 Earth radii, $R\sb{\oplus}$) from the Kepler-20 system \citep{Fressin2011}, two planet candidates (0.759 and 0.867 $R\sb{\oplus}$) from the KIC 05807616 system \citep{Charpinet2011}, and a three-planet system (0.78, 0.73, and 0.57 $R\sb{\oplus}$) orbiting KOI-961 \citep{Muirhead2012}.

The search for a second planet in the GJ 436 system began shortly after the transit detection and confirmed eccentric orbit of GJ 436b \citep{Gillon2007b,Deming2007}.  In 2008, a $\sim$5-$M\sb{\oplus}$ planet on a 5.2-day orbit was proposed by \citet[later retracted]{Ribas2008} due to three lines of evidence. 
First, the lack of detectable GJ 436b transits at the time of its 2004 discovery using radial-velocity (RV) measurements \citep{Butler2004} suggests a change in orbital inclination due to a perturber.  Second, given a circularization timescale of $\sim$30 Myr \citep{Deming2007} and the estimated 6-Gyr age of the system \citep{Torres2007}, GJ 436b's non-circular orbit suggests another planet is pumping up its eccentricity.  Third, there was evidence of a residual low-amplitude RV signal in a 2:1 mean-motion resonance with GJ 436b \citep{Ribas2008}.  The inferred planet was discredited by orbital-dynamic simulations \citep{Bean2008,Demory2009} and the absence of transit timing variations (TTVs) with two transit events with the Near Infrared Camera and Multi Object Spectrograph camera on the Hubble Space Telescope \citep{Pont2009} and over a 254-day span using ground-based H-band observations \citep{Alonso2008}.

\citet{Ballard2010a}'s analysis of 22 days of nearly continuous observations of GJ 436 during NASA's EPOXI mission ruled out transiting exoplanets $>$2.0 $R\sb{\oplus}$ outside GJ 436b's 2.64-day orbit (out to a period of 8.5 days) and $>$1.5 $R\sb{\oplus}$ interior to GJ 436b, both with a confidence of 95\%.  Aided by a $\sim$70-hour {\em Spitzer} observation of GJ 436 at 8.0 {\microns}, \citet{Ballard2010b} postulated the presence of a 0.75-$R\sb{\oplus}$ planet with a period of 2.1076 days.  However, the predicted transit was not detected in an 18-hour follow-up observation with {\em Spitzer} at 4.5 {\microns}.  The candidate transit signals in the EPOXI data were likely the result of correlated noise \citep{Ballard2010b}. 

In this paper we present {\em Spitzer} primary-transit observations of UCF-1.01 and UCF-1.02 at 4.5 {\microns} (including an independent analysis), a phase curve of GJ 436b at 8.0 {\microns} in which transits of UCF-1.01 are modeled, and a publicly-available EPOXI light curve phased to the period of UCF-1.01.  We also include secondary-eclipse observations of GJ 436b at 3.6 and 4.5 {\microns}.


In Section \ref{sec:obs}, we describe the observations and data analysis.
Section \ref{sec:tide} presents Time-series Image Denoising (TIDe, a wavelet-based technique used to improve image centers) and provides an example analysis using a fake dataset.
In Section \ref{sec:results}, we discuss the specific steps taken with each of the six {\em Spitzer} datasets, the details of our independent analysis, and transit results from the EPOXI light curve. 
Section \ref{sec:disc} describes how we eliminate false positives, our radial-velocity analysis, mass constraints on both sub-Earth-sized exoplanets, and orbital and atmospheric constraints on UCF-1.01.
Finally, we give our conclusions in Section \ref{sec:concl} and supply the full set of best-fit parameters with uncertainties in the Appendix.


\section{OBSERVATIONS AND DATA ANALYSIS}
\label{sec:obs}

\subsection{Observations}

We observed GJ 436 at 3.6 and 4.5 {\microns} using {\em Spitzer's} InfraRed Array Camera \citep{IRAC}.
Including the two previously analyzed data sets listed in Table \ref{table:ObsDates}, we present six {\em Spitzer} observations spanning just over three years.

\begin{table*}[bt]
\centering
\caption{\label{table:ObsDates} 
Observation Information}
\begin{tabular}{rccclcc}
    \hline
    \hline      
    Observation Date    & Duration  & Frame Time    & Total Frames  & {\em Spitzer} & Wavelength    & Previous      \\
                        & [minutes] & [seconds]     &               & Pipeline      & [\microns]    & Publications\tablenotemark{1}  \\
    \hline
    2008 July 14        & 4,207     & 0.4           & 588,480       & S18.18.0      & 8.0           & K10           \\
    2010 January 28     & 1,081     & 0.1           & 488,960       & S18.18.0      & 4.5           & B10           \\
    2010 June 29        & 363       & 0.4           & 49,536        & S18.18.0      & 4.5           & --           \\
    2011 January 24     & 369       & 0.4           & 51,712        & S18.18.0      & 4.5           & --           \\
    2011 February 1     & 369       & 0.1           & 168,576       & S18.18.0      & 3.6           & --           \\
    2011 July 30        & 258       & 0.4           & 36,160        & S18.18.0      & 4.5           & --           \\
    \hline
\end{tabular}
\begin{minipage}[t]{0.65\linewidth}
\tablenotetext{1}{K10 = parts were published by \citet{Knutson2010}, B10 = \citet{Ballard2010b}.}
\end{minipage}
\end{table*}

\subsection{POET Pipeline and Modeling}

Our Photometry for Orbits, Eclipses, and Transits (POET) pipeline produces systematics-corrected light curves from {\em Spitzer} Basic Calibrated Data.  We flag bad pixels, calculate image centers from a Gaussian fit, and apply interpolated aperture photometry \citep{Harrington2007} with a broad range of aperture sizes in 0.25-pixel increments.
To achieve more precise image centers in the 2010 January 28 dataset, we utilize TIDe (see Section \ref{sec:tide}).  For a more detailed description of POET, see \citet{Campo2011} and \citet{Stevenson2011}.

We model the light curve as follows:

\begin{eqnarray}
\label{eqn:full}
F(x, y, t) = F\sb{\rm s}E(t)R(t)S(t)M(x,y),
\end{eqnarray}

\noindent where \math{F(x, y, t)} is the measured flux centered at position $(x,y)$ on the detector at time $t$; \math{F\sb{\rm s}} is the (constant) system flux outside of transit events; \math{E(t)} is the primary-transit or secondary-eclipse model component; \math{R(t) = 1 - e\sp{-r\sb{0}t + r\sb{1}} + r\sb{2}(t-r\sb{3})} is the time-dependent ramp model component with free parameters $r\sb{0}-r\sb{3}$;  \math{S(t) = s\sb{0}\cos[2\pi(t-s\sb{1})/p]} is the phase variation at 8.0 {\microns} with free parameters $s\sb{0}$ and $s\sb{1}$ and $p$ being the fixed period of GJ 436b; and \math{M(x,y)} is the Bilinearly-Interpolated Subpixel Sensitivity (BLISS) map.  We follow the method described by \citet{Stevenson2011} when determining the optimal bin sizes of the BLISS maps.  

The uniform-source and small-planet equations \citep{MandelAgol2002ApJtransits} describe the secondary-eclipse and primary-transit model components.  We apply a non-linear stellar limb-darkening model \citep{Claret2000,Beaulieu2008} to UCF-1.01 transits with coefficients $a_1$-$a_4$ = (0.79660, -1.0250, 0.82228, -0.26800).  {\em Spitzer} data has well documented systematic effects that our Levenberg-Marquardt minimizer fits simultaneously with the transit/eclipse parameters.  BLISS mapping \citep{Stevenson2011} models the position-dependent systematics (such as intrapixel variability and pixelation) and linear or asymptotically constant exponential functions model the time-dependent systematics.   

A Metropolis Random-Walk Markov-chain Monte Carlo (MCMC) algorithm assesses the uncertainties \citep{Campo2011}.  Each MCMC run begins with a least-squares minimization, a rescaling of the {\em Spitzer}-supplied uncertainties so that the reduced $\chi^2=1$, and a second least-squares minimization using the new uncertainties.  We test for convergence every $10^5$ steps, terminating only when the \citet{Gelman1992} diagnostic for all free parameters has dropped to within 1\% of unity using all four quarters of the chain.  We also examine trace and autocorrelation plots of each parameter to confirm convergence visually.  We estimate the effective sample size \citep[ESS,][]{Kass1998} and autocorrelation time for each free parameter and apply the longest autocorrelation time from each event to determine the number of steps between independent samples in each MCMC chain.
We place a prior on UCF-1.01's semi-major axis ($a/R_{\star}$ = 9.1027$^{+0.0060}_{-0.0067}$) by applying its known period and GJ 436b's well constrained semi-major axis and period \citep{Knutson2011} to Kepler's third law.  Without a prior, the uncertainties for the semi-major axis (and any correlated parameters) are larger, but not unstable.  We also place a flat prior on UCF-1.02's ingress/egress time of $<0.1$ hours because it is unconstrained by the data.

\section{TIME-SERIES IMAGE DENOISING}
\label{sec:tide}

Here we describe an application of wavelets that improves image centering, resulting in more precise aperture photometry and better handling of the position-dependent systematics.  Readers primarily interested in the science results can skip to Section \ref{sec:results}.

\subsection{Introduction}


Photon noise in short exposures can cause significant shifts between the
fitted and real stellar centers.  With imprecise centering over multiple
frames, varying amounts of light fall within the improperly placed
apertures, thus increasing light-curve scatter.  The sensitivity to
precise centering increases with smaller aperture sizes, causing a given
change in aperture position to produce larger changes in flux.  To
improve centering, one could sum many exposures, but wavelet filtering
allows the same noise reduction over a shorter time span.  This is
important because \citet{Stevenson2010} detect 0.04-pixel (0.05 arcsec)
pointing variations for IRAC data over $\sim$5 seconds at \math{>10\sigma},
which limits the span of an averaging window to a few seconds.  Our
wavelet filter is called Time-series Image Denoising (TIDe, pronounced
``tidy'').  It affects only high-frequency components, such as photon
noise, without affecting low-frequency components like transits or
eclipses.  It retains the time resolution of the data.

In addition to improving aperture photometry, precise centering (see
example in Section \ref{sec:tideex}) improves our ability to model and remove
position-dependent systematics accurately, for example by reducing the
smallest meaningful bin size for BLISS mapping.  TIDe does correlate the
data in time, which complicates error analysis and makes it
computationally intense because the correlation depends on the signal
and thus varies in time.  So, we use TIDe-cleaned images only for
centering (whose uncertainties do not propagate), and perform
photometry on the unfiltered images.

As with a windowed (sliding) Fourier transform (WFT), wavelets decompose
a signal into independent contributions at each scale and location
(similar to frequency and time) within the signal.  As an example, the
Fourier transform of a piece of music can discern the average pitch and
timbre of all the instruments, but wavelets can identify individual
notes and the instruments that played them at any given time.  The
wavelet transform of a univariate time series thus has two dimensions,
for location and scale.
Unlike the WFT, wavelets do not suffer from a fixed resolution (or window
size), so they retain both good temporal resolution for high-frequency
events and good scale resolution for low-frequency events.  \citet{Torrence1998}
provide an accessible introduction to wavelets.

TIDe's improvement in precision and benefits to the light-curve fit can vary based on the source brightness, aperture size, BLISS map resolution, etc.  This method is applicable to most photon-noise-limited photometric observations where the cadence is significantly shorter than the duration or period of the time-varying object of interest.

\subsection{Description of TIDe}


 
TIDe applies discrete wavelet denoising independently to multiple time series, each comprised of the values measured in a single pixel as a function of time (i.e., frame number).  
Every pixel is associated with such a time series, and each one is denoised independently of adjacent image pixels.  
The transformed data (known as wavelet coefficients) for each pixel time series have a location (or time) dimension and a scale (or level) dimension.  The wavelet coefficients map the discrete wavelet to the data at each scale and instant in time. 
The lowest level (or finest scale) of decomposition in a wavelet transform describes how the data change on the shortest timescales.  
Assuming that this level is dominated by noise, we can eliminate wavelet coefficients with magnitudes below a certain threshold (hard thresholding) or merely attenuate them (soft thresholding) to reduce their contribution to the overall signal.  Adjusting a collection of estimates together in this way can be shown to improve the average quality of the estimates by introducing a small bias that is more than compensated for by reduced variance.
These techniques can also be applied to successively higher levels, but they have less impact at longer timescales where the signal dominates over the noise.  
After thresholding, we recombine all of the adjusted wavelets to generate a less-noisy version of the original pixel time series.
For each frame, an image is re-created from the many denoised time series, and centering is performed using that image.  There is no explicit spatial denoising, but to the extent that there are spatial correlations between images at different epochs, there is an implicit spatial denoising in the processed image that improves center estimation.
The effectiveness of TIDe is determined by the threshold at which wavelet coefficients are zeroed, the type of thresholding technique applied, and the number of levels to which the method is applied \citep{Donoho1994, Chang2000}.


Various wavelet thresholding techniques exist, each with its own advantages and disadvantages.  Two common methods for suppressing noise are hard and soft universal thresholding and are defined, respectively, as follows:

\begin{equation}
\omega = y I(|y|>T),
\end{equation}
\begin{equation}
\omega = {\rm sgn}(y) (|y|-T) I(|y|>T),
\end{equation}

\noindent where $y$ ($\omega$) are the original (denoised) wavelet coefficients at a particular level,  $I$ is the Indicator function (1 if true, 0 if false), and $T$ is some threshold limit.  In both instances, if a particular wavelet coefficient, $y$\sb{i}, is less than $T$, then $\omega$\sb{i} = 0.  With hard thresholding, the remaining coefficients are unaltered; however, soft thresholding shrinks these coefficients by the threshold limit.

There are many ways to estimate the value of $T$, including VisuShrink, SURE Shrink, and Bayes Shrink \citep{Chang2000}.  With TIDe, we implement the last technique because it establishes a thresholding rule that is optimal in terms of minimizing the expected RMS error in the denoised time series under flexible assumptions for the true time series signal (i.e., it minimizes the Bayes Risk for a squared-error loss function).
Bayes Shrink employs soft thresholding because its optimal estimator yields a smaller risk than hard-thresholding's estimator.  
The optimal threshold value is determined as follows \citep[see][]{Chang2000}.  In some instances, the noise variance, $\sigma^2$, may be known {\em a priori}.  If this is not the case, it is estimated from the robust median estimator \citep{Donoho1994}:

\begin{equation}
\sigma = \frac{{\rm Median}(|Y\sb{1}(y)|)}{0.6745},
\end{equation}

\noindent where $Y\sb{1}(y)$ represents the wavelet coefficients, $y$, at the lowest level (or finest scale) of decomposition.  Next, we estimate the variance of $Y(y)$ at a particular level $j$, assuming zero mean, by:

\begin{equation}
\sigma_{j}^{2} = \frac{1}{n}\sum_{i=1}^{n}Y_{j}^{2}(y_{i}).
\end{equation}

\noindent where $n$ is the number of wavelet coefficients at that level.  Our observation model (data = signal + noise) tells us that $\sigma_{j}^{2} = \sigma_{x}^{2} + \sigma^{2}$, where $\sigma_{x}^{2}$ is the signal variance.  To account for the case where $\sigma^{2} > \sigma_{j}^{2}$, we calculate $\sigma_{x}$ as follows:

\begin{equation}
\sigma_{x} = \sqrt{{\rm max}(\sigma_{j}^{2}-\sigma^{2},0)}.
\end{equation}

\noindent Finally, the optimal threshold at a particular level is determined by:

\begin{equation}
T_{j} = \frac{\sigma^{2}}{\sigma_{x}}.
\end{equation}

\noindent In the event that $\sigma^{2} > \sigma_{j}^{2}$ ($T_{j} = \infty$), all of the wavelet coefficients are set to zero.  

In this paper, we use the Biorthogonal 5.5 discrete wavelet from the PyWavelets package to apply soft thresholding with Bayes Shrink to the designated scale levels.

\subsection{TIDe Example}
\label{sec:tideex}

\begin{figure}[ht]
\centerline{
\includegraphics[clip,width=1.0\columnwidth]{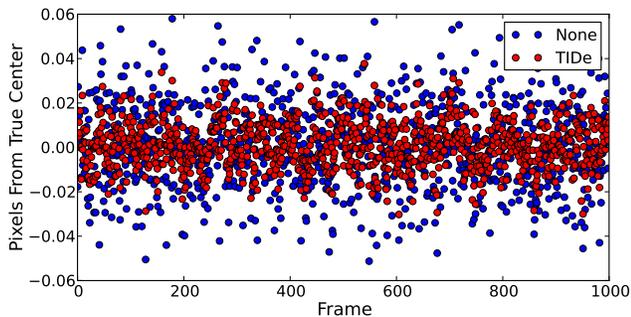}
}
\caption{\label{fig:denoising}{\small
Illustrative example of TIDe that compares image centers from simulated noisy frames (blue) to their denoised counterparts (red).  In each case, only the $y$ component of the position is plotted relative to the true center.  In this typical example, the standard deviation in the pointing (a measure of precision) decreased from 0.019 to 0.011 pixels.
}}
\end{figure}

We generate a series of 1000 test frames, each containing a 2D Gaussian with a width of 0.7 pixels, a peak flux of 1000, and centered at (4.5, 4.5) in a 10$\times$10 frame with the lower-left corner indexed as (0, 0).  We then added a background flux offset of 100 and applied Poisson noise to each frame.  We performed 2D Gaussian centering to derive the blue points plotted in Figure \ref{fig:denoising}.  For the points in red, we applied TIDe to the frames using a Biorthogonal 5.5 discrete wavelet (from the PyWavelets package) then recalculated the image centers with the same 2D Gaussian centering routine.  In each case, only the $y$ component of the position is plotted for each frame.  Using TIDe, the standard deviation in the pointing about the true center decreased from 0.019 to 0.011 pixels, for a typical improvement of $\sim$40\%.  We see even better results with the 2010 January 28 data set, where TIDe improved the pointing precision by $\sim$70\%.


\section{LIGHT-CURVE FITS AND RESULTS}
\label{sec:results}

\begin{figure}[ht]
\centering{
\includegraphics[clip,width=1.0\columnwidth]{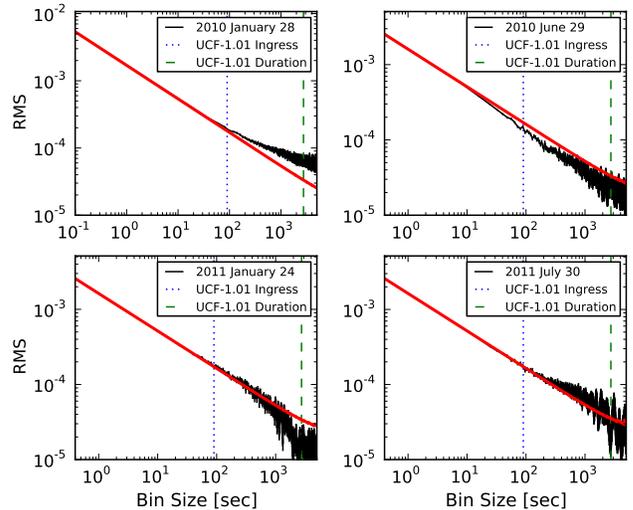}
}
\caption{\label{fig:rms}
Normalized RMS residual flux \vs bin size (in black) for four 4.5-{\micron} light curves.  Black vertical lines at each bin size depict 1$\sigma$ uncertainties on the RMS residuals ($RMS / \sqrt{2N}$, where $N$ is the number of bins).  The red lines show the predicted standard error for Gaussian noise. 
The dotted and dashed lines indicate the scale length of UCF-1.01's best-fit transit ingress and duration times, respectively.  The excess RMS above the red line in the top left panel indicates correlated noise at timescales near UCF-1.01's transit duration and is discussed in Section \ref{sec:cp21}.
}
\end{figure}

\begin{figure}[ht]
\centering{
\includegraphics[width=1.0\columnwidth, clip]{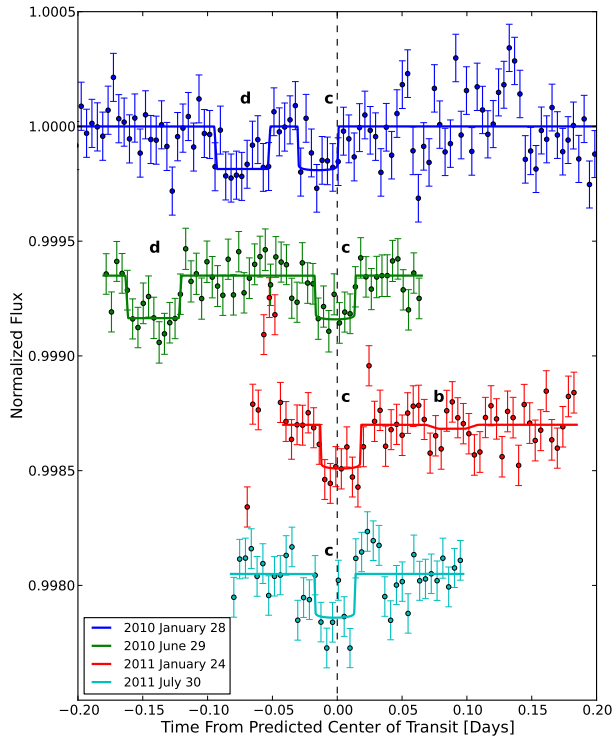}
}
\caption{\label{fig:lc}
Four 4.5-{\micron} {\em Spitzer} light curves of GJ 436 with best-fit models.  The flux values are corrected for systematics, normalized to the system brightness, and binned (with 1$\sigma$ error bars).  Light curves are vertically separated for ease of comparison.  
The single GJ 436b eclipse, four UCF-1.01 transits, and two UCF-1.02 candidate transits are indicated by the letters b, c, and d, respectively.  The transits distinguish themselves by their consistency in depth and duration.  Although UCF-1.01's 2010 January 28 best-fit transit time is 20 {\pm} 7 minutes earlier than our predicted time (dashed line), the parameter's probability distribution is bimodal (see Figure \ref{fig:cp21corr}) and the other peak is only 6 {\pm} 7 minutes early.  We quote the median transit time in Table \ref{tab:aei} to encompass both possibilities.
The three remaining UCF-1.01 transit times (see Table \ref{tab:aei}) occur within five minutes of the predicted times and have a typical uncertainty of {\pm}3 minutes.  The episodic scatter in flux is most likely due to stellar activity, which is expected for an M dwarf and seen in many observations of this system.
Using the non-detection of GJ 436b in the 2011 January 24 dataset, we place a 3$\sigma$ upper limit of 95 ppm on its eclipse depth, resulting in a brightness temperature $<$780 K.  This new 4.5-{\micron} secondary-eclipse observation supports a methane-deficient and carbon monoxide-rich dayside atmosphere \citep{Stevenson2010}.
}
\end{figure}

We present the scaling of the RMS model residuals {\vs} bin size (a test of correlation in time) in Figure \ref{fig:rms} for all four 4.5-{\micron} {\em Spitzer} observations.
Figure \ref{fig:lc} displays our reanalysis of GJ 436 data \citep{Ballard2010b} plus three new {\em Spitzer} light curves at 4.5 {\microns}.  Two fortuitous detections of UCF-1.01 appeared during atmospheric characterization observations of GJ 436b \citep{Stevenson2010}.  Using these data and a tentative detection at 8.0 {\microns} (see Section \ref{sec:bo41}) to estimate its orbital period, we extrapolated UCF-1.01 transit times forward by six months to predict an event (2011 July 30) during the next observing window.
We supply correlation plots and histograms in Appendix \ref{sec:corr} and the full set of best-fit parameters from a 2.4$\times$10\sp{6}-iteration joint fit in Appendix \ref{sec:params}.
Below, we discuss each observation in detail to explain how we arrived at the final results.

\subsection{2010 January 28 (4.5-{\micron} Spitzer Observation)}
\label{sec:cp21}

{\em Spitzer} program 541 (Sarah Ballard, P.I.) monitored GJ 436 continuously for $\sim$18 hours using 0.1-second exposures.  The short exposure time allows us to apply TIDe to the lowest four levels (L4) of wavelet decomposition (see Section \ref{sec:tide}), resulting in a maximum affected time resolution of 1.6 seconds.  In \citet{Stevenson2010}, we detect {\em Spitzer} pointing changes on timescales as short as $\sim$5 seconds, longer than TIDe's timescale.
In calculating image centers with and without TIDe, we find that the position consistency between consecutive denoised frames improves by more than a factor of three, resulting in an RMS of 0.0015 pixels in $x$ and 0.0011 pixels in $y$.  More precise image centers decrease flux scatter with smaller aperture sizes and aid the BLISS map in modeling the position-dependent systematics.  We apply $5\times$-interpolated aperture photometry to the unmodified frames to avoid the computationally prohibitive calculation of estimating uncertainties for the denoised frames, which are correlated in time.  

\begin{figure}[!h]
\centering
\includegraphics[width=\columnwidth,clip]{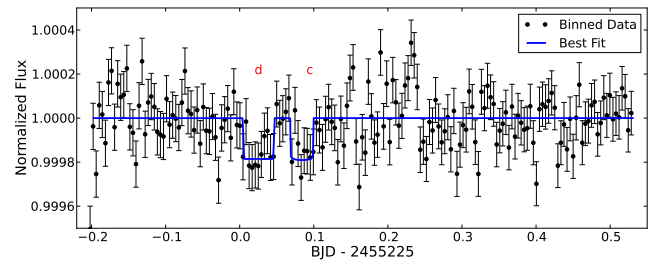}
\caption{\label{fig:gj436cp21}{\small
Full light curve from 2010 January 28 {\em Spitzer} observation with transits of UCF-1.01 (right) and UCF-1.02 (left).  The flux values are corrected for systematics, normalized to the system brightness, and binned (with 1$\sigma$ error bars).  The solid line depicts the best-fit model.  Excluding the excess flux near 2455225.23 in the model fit does not significantly alter the best-fit solution.
}}
\end{figure}

Photometry generates consistent transit depths for all tested apertures from 1.25 to 4.50 pixels, but an aperture size of 2.25 pixels produces the lowest standard deviation of the normalized residuals (SDNR).  We estimate the background flux using an annulus from 7 to 15 pixels centered on the star.
The light curve (see Figure \ref{fig:gj436cp21}) exhibits a strong initial increase in pixel sensitivity that we do not model (preclip, $q$ = 10,000).  As with B10, we detect excess flux (possibly due to stellar activity) near BJD 2455225.23 that contributes to the observed correlated noise in Figure \ref{fig:rms}.
We note that frames 19,780 -- 19,839, 82,180 -- 82,239, 165,380 -- 165,439, 419,780 -- 419,839, and 451,780 -- 451,839 are shifted horizontally by one pixel, so we flag these frames as bad.
A probable micro-meteor impact at BJD $\sim$2,455,224.976 caused a sudden shift in pointing before returning to its original location.  Simultaneously, the background scatter increased by $\sim$50\% and remained elevated until the end of the observation.  
We apply a BLISS map bin size ($x$, $y$) of 0.007 $\times$ 0.005 pixels and set the minimum number of points per bin to six to disregard the observed excursion.

\subsection{2010 June 29 (4.5-{\micron} Spitzer Observation)}

\begin{figure}[h]
\centering
\includegraphics[width=\columnwidth,clip]{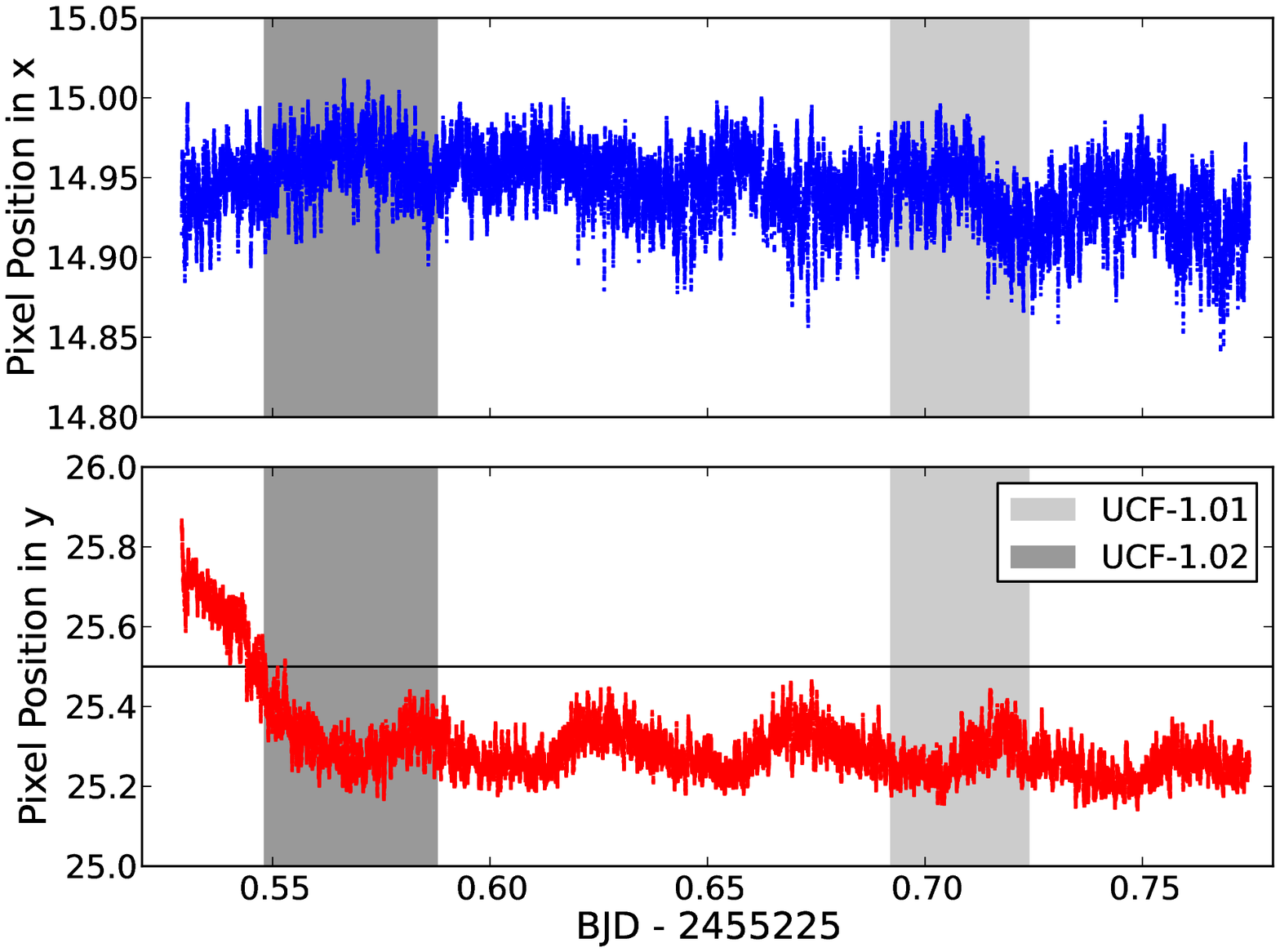}
\includegraphics[width=\columnwidth,clip]{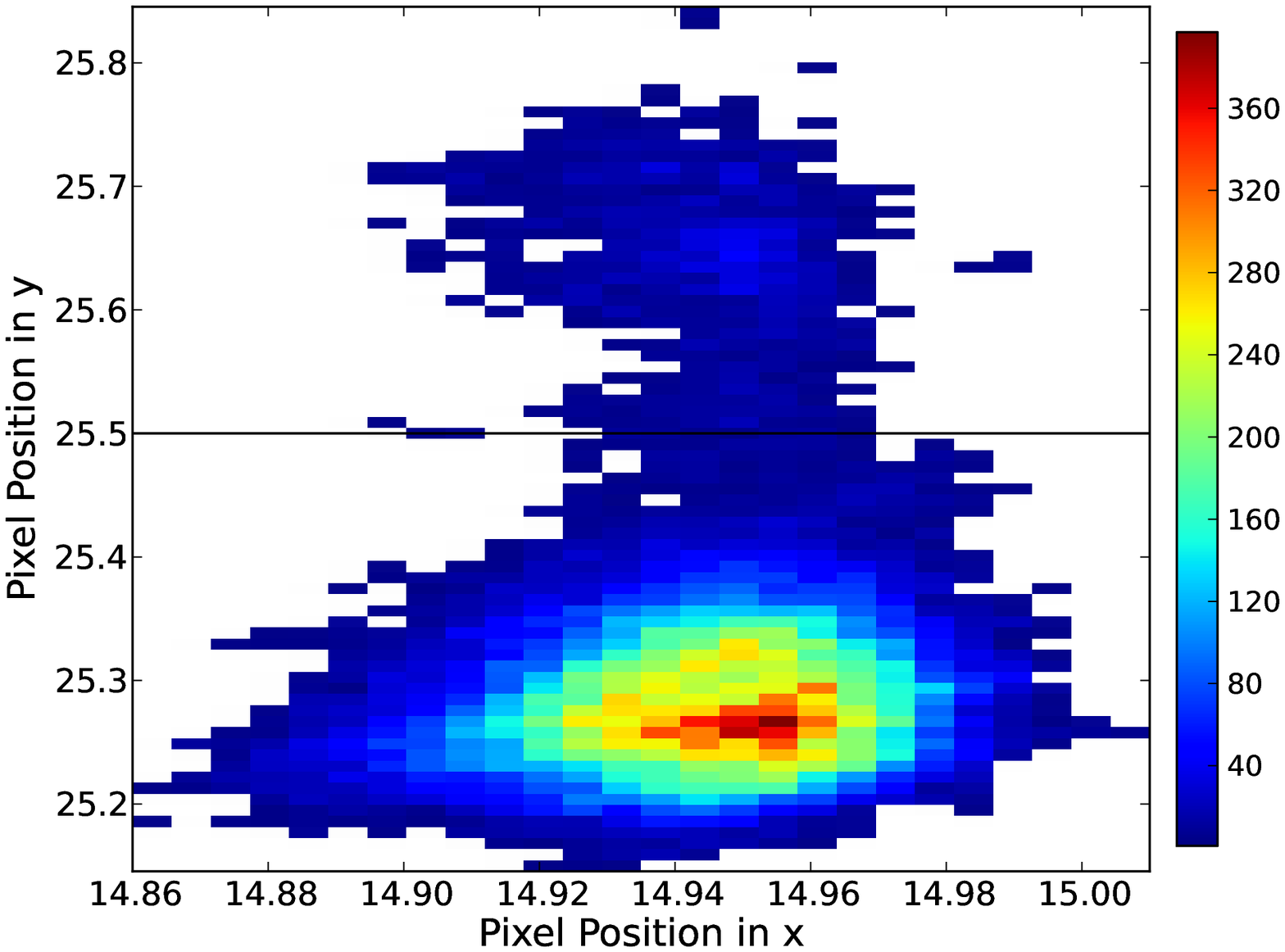}
\caption{\label{fig:gj436bs22}{\small
Image centers {\vs} time (upper 2 panels) and pointing histogram (lower panel, number of image centers within a given bin) for the 2010 June 29 dataset.  The times during transit are shaded in gray.  Initial telescope drift hampers our ability to effectively model position-dependent systematics during and prior to the UCF-1.02 transit.
}}
\end{figure}

Our {\em Spitzer} program 60003 (Joseph Harrington, P.I.) monitored GJ 436 for six hours using 0.4-second exposures.  The mean image center is located at pixel (15, 25), near the top of the 32$\times$32 array, thus restricting aperture sizes to $\le$5.50 pixels.  
Using a background sky annulus from 10 to 30 pixels, we find that the lowest SDNR occurs with a 5$\times$-interpolated aperture 5.00 pixels in radius.  The BLISS map uses bins of size 0.006 $\times$ 0.009 pixels and with at least four points per bin.
We test image centers generated from L3 TIDe (3.2-second maximum time resolution) but find no improvement in the SDNR.  This is likely due to the smaller improvement in image centers and significantly larger aperture size, relative to the 2010 January 28 dataset.  The final analysis did not use TIDe.
For this dataset, the telescope pointing does not stabilize until midway through the transit of UCF-1.02 (see Figure \ref{fig:gj436bs22}).  As a result, the position-dependent systematic is poorly constrained during and prior to the transit.  This may be the source of UCF-1.02's variable transit duration, which decreases with smaller aperture sizes.  More observations are necessary to confirm its parameters.

\subsection{2011 January 24 (4.5-{\micron} Spitzer Observation)}

{\em Spitzer} program 60003 performed a second six-hour observation of GJ 436 with 0.4-second exposures.
We find that 10$\times$-interpolated aperture photometry outperforms 5$\times$-interpolated and minimizes SDNR with an aperture size of 5.25 pixels and a background sky annulus from 10 to 30 pixels.  We flag 54 frames (28,426 -- 28,479) as bad due to a one-pixel horizontal shift, as observed previously in a dataset above.
We clip the first 6,000 observations due to a strong increase in flux, possibly due to stellar activity (see Figure \ref{fig:lc}).
Near 2455585.771, we observe a sudden shift in the telescope pointing that, again, correlates with an increase in background noise.
To remove this excursion from our models, the BLISS map uses bins with eight or more points and a size of 0.016 $\times$ 0.008 pixels.  As with the previous dataset and for the same reasons, TIDe centers have little effect on the resulting photometric light curve.

\subsection{2011 February 1 (3.6-{\micron} Spitzer Observation)}

\begin{figure}[!h]
\centering
\includegraphics[width=\columnwidth,clip]{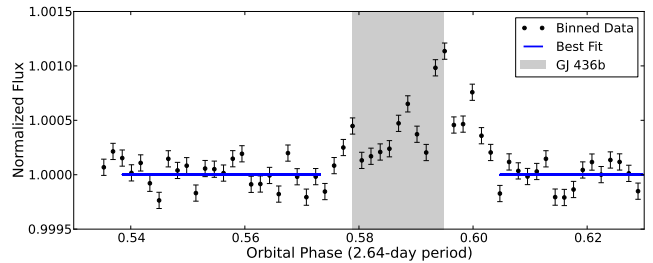}
\caption{\label{fig:gj436bs13}{\small
{\em Spitzer} 3.6-{\micron} light curve from 2011 February 1 with GJ 436b's time of secondary eclipse shaded in gray.  The flux values are corrected for systematics, normalized to the system brightness, and binned (with 1$\sigma$ error bars).  The solid line depicts the best-fit baseline model.  Stellar activity prohibits us from fitting the eclipse and measuring its depth.  However, by visually comparing the binned points within the shaded region to those immediately outside, the data appear to be consistent with a relatively deep eclipse depth \citep{Stevenson2010}.
}}
\end{figure}

Our {\em Spitzer} program 60003 also observed GJ 436 at 3.6 {\microns} with 0.1-second exposures.  We apply 5$\times$-interpolated aperture photometry with an aperture size of 2.75 pixels and a background sky annulus from 7 to 15 pixels.  We clip the first 10,000 frames due to a steep ramp and frames 70,000 -- 125,000 due to suspected stellar activity (see Figure \ref{fig:gj436bs13}).  Because GJ 436b's time of secondary eclipse occurs during the period of increased stellar activity, we do not fit the eclipse or calculate uncertainties.

\subsection{2011 July 30 (4.5-{\micron} Spitzer Observation)}

{\em Spitzer} monitored GJ 436 for 4.3 hours using 0.4-second exposures (program 70084, Joseph Harrington, P.I.).
Photometry generates consistent transit depths for apertures between 1.75 and 6.00 pixels.  The final run applies 10$\times$-interpolated, 5.00-pixel aperture photometry with a background sky annulus from 10 to 30 pixels.
During these observations, the telescope pointing experiences two deviations, at BJD 2,455,772.766 and 2,455,772.870.  The background variance increases with the first event but slightly decreases with the second event.  
BLISS mapping utilizes a bin size of 0.012 $\times$ 0.006 pixels with a minimum of six points per bin to exclude points from either excursion.  Again, TIDe centers have little effect on the resulting photometric light curve.

\subsection{2008 July 14 (8.0-{\micron} Spitzer Observation)}
\label{sec:bo41}

\begin{figure}[ht]
\centering
\includegraphics[width=\columnwidth]{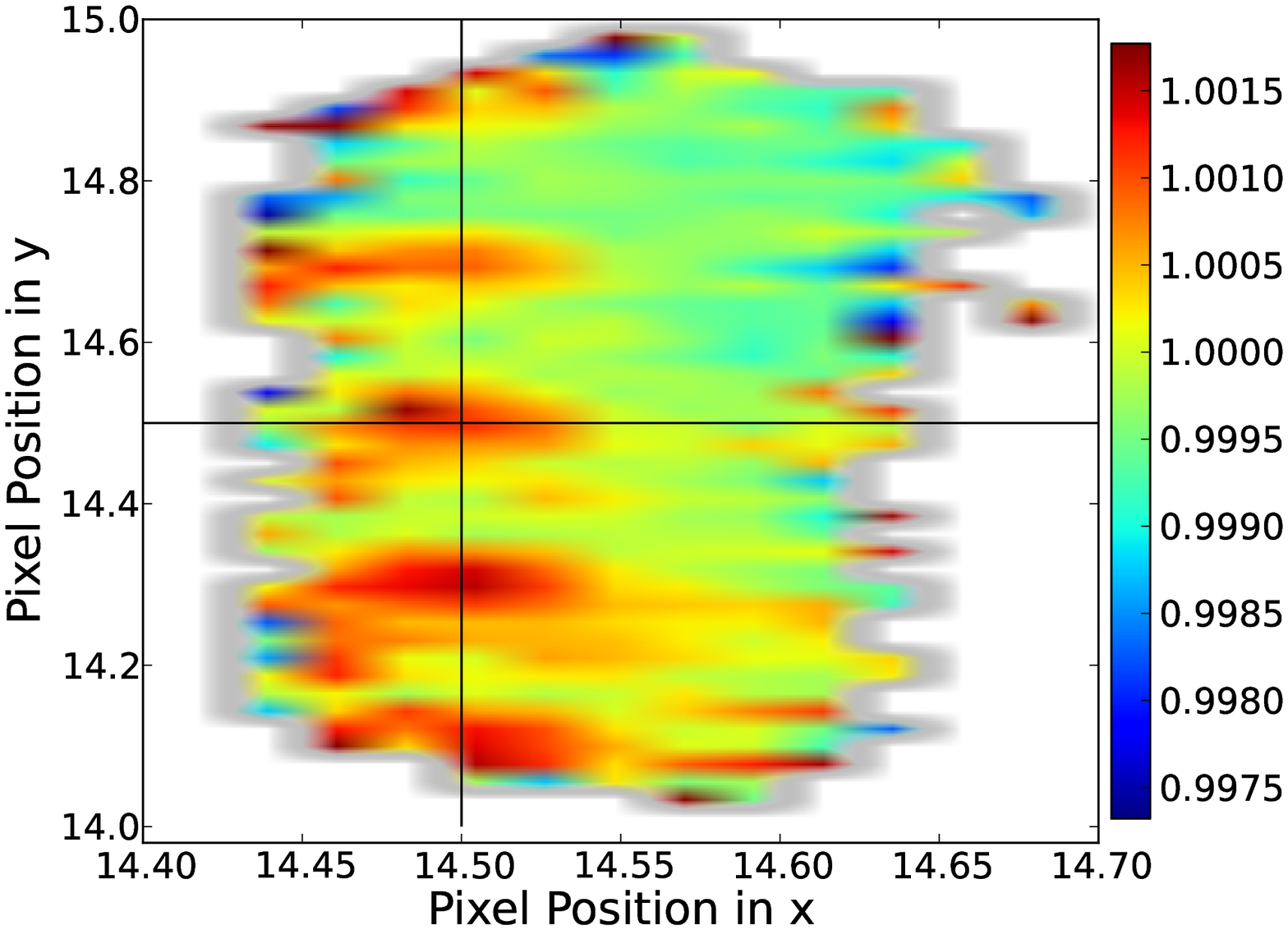}
\includegraphics[width=\columnwidth]{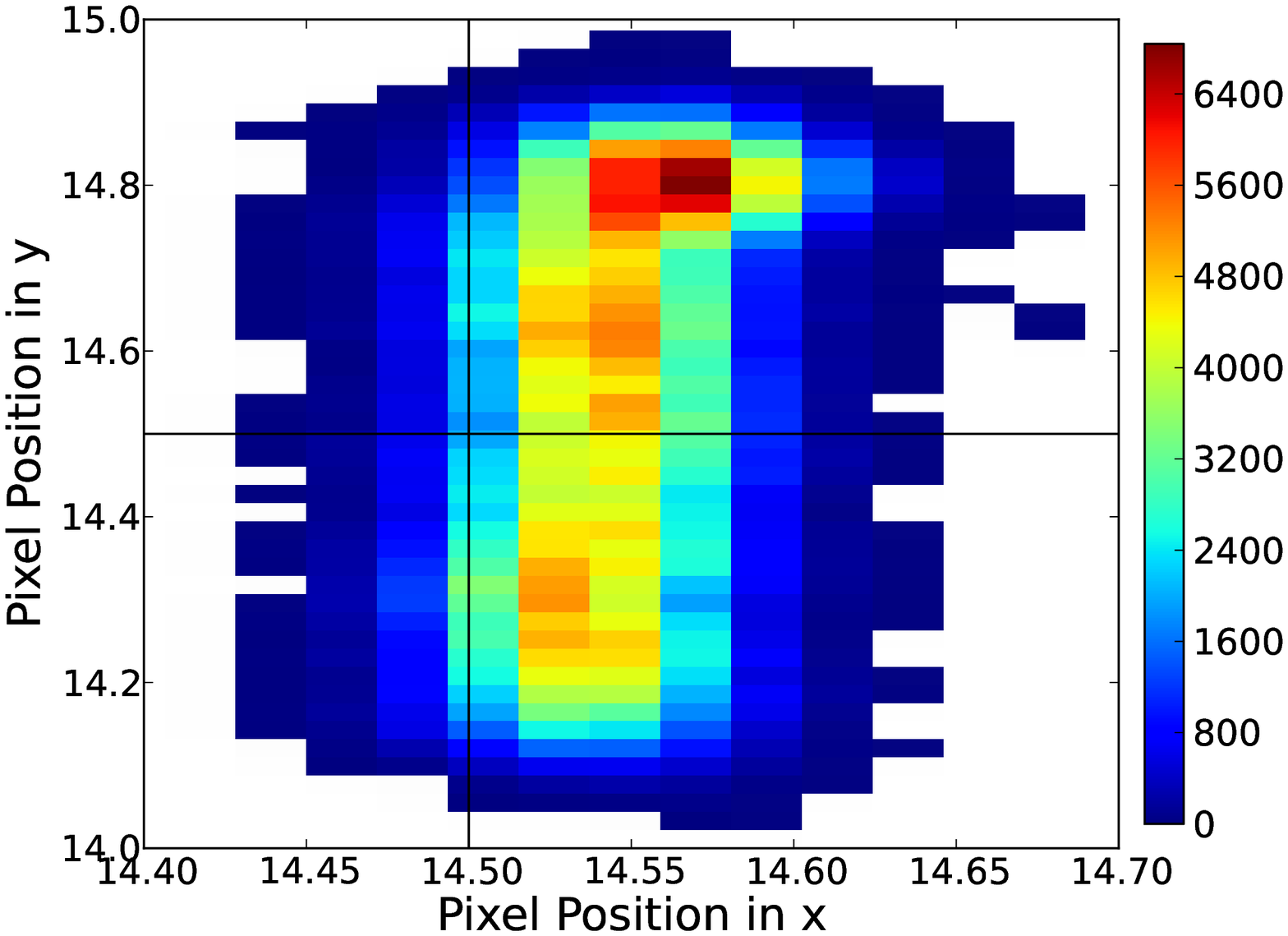}
\caption{\label{fig:bliss}{\small
BLISS map (top) and pointing histogram (bottom) for the 2008 July 14 dataset.  Pixelation, a position-dependent systematic, is depicted by the colors in the BLISS map, where redder (bluer) colors indicate more (less) flux within the aperture.  Peaks repeat every 0.2 pixels because we applied 5$\times$-interpolated aperture photometry.  Smaller interpolation factors result in larger spacing between peaks but also a stronger systematic between peaks.
The horizontal and vertical black lines depict pixel boundaries.
}}
\end{figure}

\begin{figure}[!h]
\centering
\includegraphics[width=\columnwidth,clip]{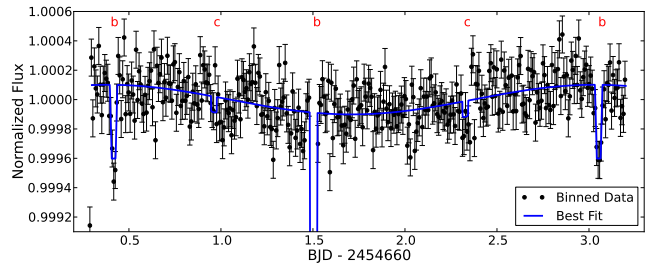}
\caption{\label{fig:gj436bo41}{\small
{\em Spitzer} light curve of GJ 436 at 8.0 {\microns}.  The flux values are corrected for systematics, normalized to the median system brightness, and binned (with 1$\sigma$ error bars).  The solid blue line depicts the best-fit model.
The light curve contains two eclipses (first and last events) and one transit (center event) of GJ 436b and two transits (second and fourth events) of UCF-1.01.  
The two UCF-1.01 transit depths have a combined significance of 2$\sigma$, which is insufficient to claim a detection, but we used the timing of the latter to predict the 2001 July 30 transit.
The difference in brightness temperatures between GJ 436b's dayside and nightside causes the observed sinusoidal variation in the light curve.
The peak-to-peak flux difference is 200 {\pm} 50 ppm (4$\sigma$ significance).  This corresponds to a brightness temperature difference of 110 {\pm} 60 K, which favors a relatively efficient dayside-to-nightside energy redistribution.  The peak flux is shifted by 0.7 {\pm} 4.6 hours prior to secondary eclipse.
}}
\end{figure}

{\em Spitzer} program 50056 (Heather Knutson, P.I.) observed GJ 436 for $\sim$70 hours from 2008 July 12 to 2008 July 15.
At the best aperture size of 3.75 pixels (and a background sky annulus from 7 to 15 pixels), we find that the light curve exhibits a measurable position-dependent systematic, identified as pixelation \citep{Stevenson2011}.  The BLISS map fits and removes pixelation (see Figure \ref{fig:bliss}) using a bin size of 0.022 $\times$ 0.022 pixels and at least four points per bin.  We model the initial time-dependent ramp using an asymptotically constant exponential function after clipping the first 3,000 frames.  A sinusoidal function with a linear correction fits the phase variation of GJ 436b (see Figure \ref{fig:gj436bo41}).  We set a prior on the inclination and semi-major axis of UCF-1.01 using the best-fit results from the 4.5-{\micron} joint fit.
The UCF-1.01 transit at BJD 2,454,662.328 is the same candidate transit reported by \citet{Ballard2010b} using a $\sim$2.1-day period estimated from EPOXI observations. Their Figure 5 incorrectly reports the BJD.  We used the timing of this transit to successfully predict the 2001 July 30 transit.  The best-fit radius ratio from both transit events in this light curve is 0.010 {\pm} 0.003, which is consistent with the best-fit result using the four 4.5-{\micron} light curves.

\subsection{Independent Analysis}

We sought an independent analysis to confirm our results. Nikole Lewis analyzed each of the 4.5-{\micron} datasets without knowing the times or depths of the transits. In addition to using her own photometric pipeline, she applied a new pixel-mapping routine that shares a heritage with the method from \citet{Ballard2010b}.
This new pixel-mapping method was developed to recover the relative
flux variations as a function of orbital phase from the {\it Spitzer}
3.6-{\micron} and 4.5-{\micron} full orbit light curves of HD~189733b,
HD~209458b, HAT-P-2b, and HAT-P-7b (PI:Knutson; PID 60021).  Similar to the
BLISS method, the pixel-mapping technique developed by
Lewis uses nearest neighbors to calculate flux as a function of
position on the detector, but in her method the distances are weighted
according to a Gaussian distribution.  In addition to stellar centroid positions, Lewis makes use
of the ``noise pixel'' parameter given in frame headers to determine the
nearest-neighbors to a given data point (Lewis {\em et al.}, in prep.).  
This routine improves on Ballard's method by calculating the pixel map at each iteration without being computationally prohibitive.
 
\begin{figure}
\centering
  \includegraphics[width=\columnwidth,clip]{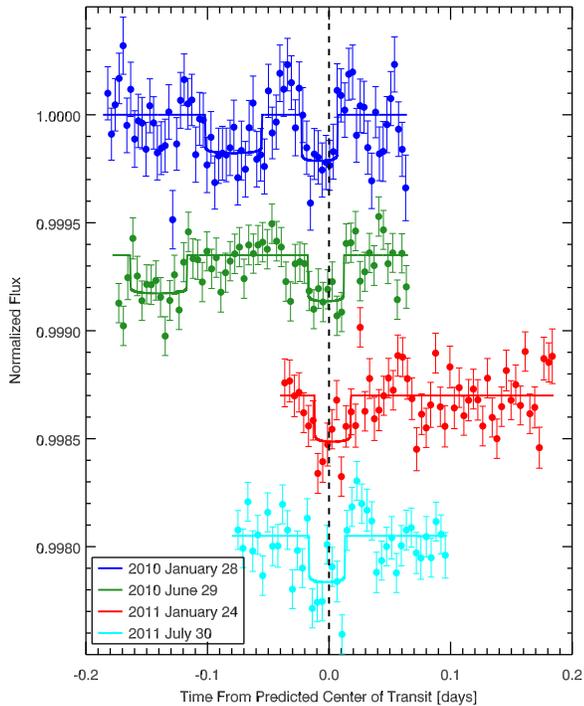}\\
  \caption{\label{lewis_lc}{\small
Four 4.5-{\micron} {\em Spitzer} light curves of GJ~436 with best-fit models from an independent analysis by Lewis.
She corrects flux measurements for intrapixel sensitivity variations using a pixel-mapping technique and for the presence of the well documented {\it Spitzer} time-dependent systematic (ramp).  
A fixed-width symmetric Gaussian fits centroid positions in the region near the brightest pixel in each subarray frame.  The best photometric aperture size is 2.25 pixels for the 2010 January 28 dataset and 5.0 pixels for the other datasets.  The non-linear limb-darkening coefficients for GJ~436 are those from \citet{Knutson2011}. After the location of the transit(s) in each dataset were identified individually, Lewis performed a simultaneous fit between all four datasets using a Levenberg-Marquardt minimization scheme.  A MCMC algorithm determined the uncertainty in the fit parameters as well as identified other possible solutions.  The goal of this analysis was to confirm the presence and shape of transit(s) in each dataset.  Improvements to treatment of the systematics in these observations is possible, but they are unlikely to significantly change the estimated planetary parameters.}}
\end{figure}

Pixel mapping is essential to detecting the weak transit signals in
these data.  For example, the 2010 January 28 dataset requires an accurate pixel mapping routine, at minimum, to detect UCF-1.01 and benefits from more precise image centers with TIDe to more clearly distinguish UCF-1.02. We have found that without a pixel-mapping routine, one cannot reproduce all of the observed transits.
Lewis uncovered transits of UCF-1.01 in the 2010 June 29, 2011 January 24, and 2011 July 30 datasets with ease and, once informed of the additional planet, identified both UCF-1.02 transits and the remaining UCF-1.01 transit (see Figure \ref{lewis_lc}).  Her final transit times, depths, and durations for both planets are all within 1.5$\sigma$ of our best-fit results.

\subsection{EPOXI Observation}

\begin{figure}[!h]
\centering
\includegraphics[width=\columnwidth,clip]{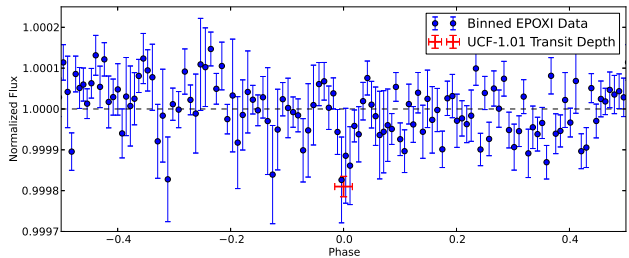}
\caption{\label{fig:epoxi}{\small
EPOXI light curve phased to the period of UCF-1.01 using the best-fit period and nearest ephemeris time (2008 July 14 dataset).  Blue circles represent the binned EPOXI data with 1$\sigma$ uncertainties.  The red cross depicts the duration and depth (with a 1$\sigma$ uncertainty) of UCF-1.01's transit.  The EPOXI data are consistent with a UCF-1.01 transit.
}}
\end{figure}

NASA's EPOXI mission observed GJ 436 nearly continuously during 2008 May 5 -- 29 \citep{Ballard2010a}.  The light-curve data are available from EPOXI's archive.  After masking the transits of GJ 436b, we divide the light curve by the median flux value, phase it according to the best-fit UCF-1.01 period (see Table \ref{tab:aei}), and bin the results.  Figure \ref{fig:epoxi} illustrates a visible decrease in the observed flux at the correct phase that is consistent with the transit depth and duration of UCF-1.01 derived from the {\em Spitzer} data.  The quality of the light curve is such that the data neither prove nor disprove the existence of UCF-1.01.

\section{DISCUSSION}
\label{sec:disc}

Without continuous monitoring of GJ 436 for two consecutive transits at the most photometrically-precise wavelengths (3.6 and 4.5 {\microns}), we isolate the true period from integer multiples or whole number fractions by other means.  Integer multiples (i.e., 2, 3, 4...) of the orbital period (see Table \ref{tab:aei}) cannot account for all of the observed transits; whole number fractions (i.e., 1/2, 1/3, 1/4...) are eliminated by investigating the bevy of available GJ 436 {\em Spitzer} data at predicted transit times.  We find evidence against periods of $\sim$0.6829 and $\sim$0.4553 days by non-detections of UCF-1.01 in a 2008 January 30 observation at 3.6 {\microns} and a 2008 June 11 observation at 8.0 {\microns}, respectively.  The single UCF-1.01 detection in the 2010 January 28, 18-hour observation dismisses even shorter periods.

\begin{table}
\centering
\caption{\label{tab:aei} 
Transit model best-fit values and other parameters}
\begin{tabular}{@{}rcc@{}}
    \hline
    \hline
    Parameters              & UCF-1.01                       & UCF-1.02 \\
    \hline
    $R\sb{p}/R\sb{\star}$   & 0.0138 {\pm} 0.0009$^{a}$     & 0.0136 {\pm} 0.0012       \\
    $i$ [$\sp{\circ}$]      & 85.17$^{+0.8}_{-0.16}$$^{a}$  & --                        \\
    $a/R\sb{\star}$         & 9.10 {\pm} 0.07$^{a}$         & --                        \\ 
    Impact Parameter        & 0.77$^{+0.02}_{-0.15}$        & --                        \\
    Transit depth [ppm]     & 190 {\pm} 25                  & 186 {\pm} 30$^{a}$        \\
    Duration [t\sb{4-1}, hr] 
                            & 0.76$^{+0.15}_{-0.03}$        & 1.04$^{+0.26}_{-0.15}$$^{a}$  \\
    Ingress/Egress [hr]     & 0.025$^{+0.002}_{-0.004}$     & $<$0.1$^{a}$              \\
    Transit Times [MJD$_{TDB}$]$^{b}$
                            & 5225.090$^{+0.004}_{-0.005}$$^{c}$  
                                                            & 5225.026 {\pm} 0.003$^{a}$    \\
                            & 5376.7078$^{+0.0014}_{-0.0021}$$^{a}$  
                                                            & 5376.568$^{+0.003}_{-0.007}$$^{a}$    \\
                            & 5585.6889$^{+0.0020}_{-0.0018}$$^{a}$  & --                        \\
                            & 5772.8069$^{+0.0009}_{-0.0029}$$^{a}$  & --                        \\
    Mean Period [Days]      & 1.365862 {\pm} 8$\times$10$^{-6}$      & --                        \\
    Ephemeris [MJD$_{TDB}$]$^{b}$ 
                            & 5772.8086 {\pm} 0.0016        & --                        \\
    Radius [$R\sb{\oplus}$] & 0.66 {\pm} 0.04               & 0.65 {\pm} 0.06           \\
    Mass [$M\sb{\oplus}$]$^{d}$
                            & 0.28 {\pm} 0.06               & 0.27 {\pm} 0.07           \\
    \hline
\end{tabular}
\begin{minipage}[t]{0.90\linewidth}
$^{a}$ Fitted values. \\
$^{b}$ MJD = BJD - 2,450,000. \\
$^{c}$ We choose the median value because the distribution is bimodal. \\
$^{d}$ Assuming an Earth-like density of 5.515 g/cm$^{3}$.
\end{minipage}
\end{table}

\subsection{Eliminating False Positives}
\label{sec:falsepositives}

\begin{figure}[h]
\centering
\includegraphics[width=\columnwidth,clip]{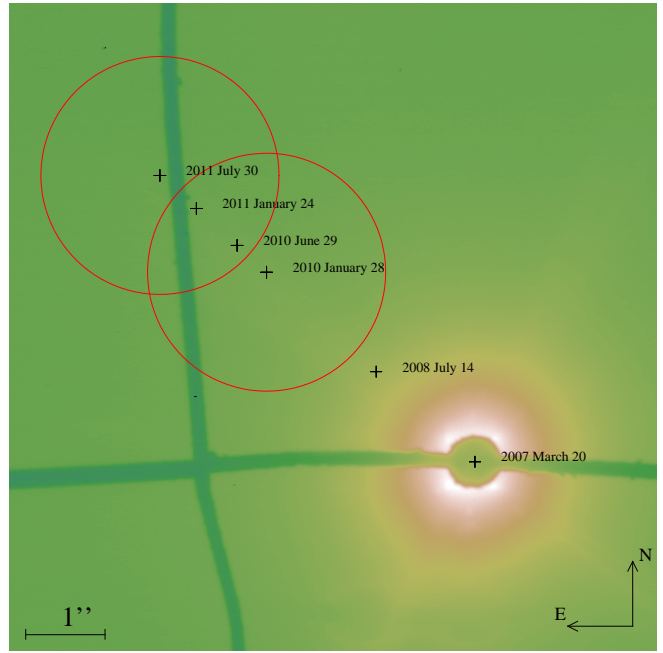}
\caption{\label{fig:vlt}
Very Large Telescope H-band observation on 2007 March 20 using the NAOS-CONICA instrument with adaptive optics \citep{Montagnier2011}.
We search for faint background systems by blocking the light from GJ 436 using a 0.7{\arcsec} Lyot coronagraphic mask.  The dark green lines are mask support wires.  The ``+'' symbols indicate the position of the GJ 436 system for this observation and at each transit epoch of UCF-1.01.  Red circles indicate the minimum photometric aperture size (1.25 pixels) for which transit signals from the first and last confirmed events may still be clearly distinguished against the background noise.  If a background system were the source of the transit-like events, it must put light in the overlapping region.  To produce the observed transit depth, the hypothetical system must be no more than 9.3 magnitudes fainter than GJ 436, assuming a total eclipse of one of the objects.  We eliminate objects brighter than $\Delta$H = 12.7 relative to GJ 436 with a confidence of 5$\sigma$.  
There are also no objects listed in any catalog within this region.
}
\end{figure}

\begin{figure}[h]
\centering
\includegraphics[width=\columnwidth,clip]{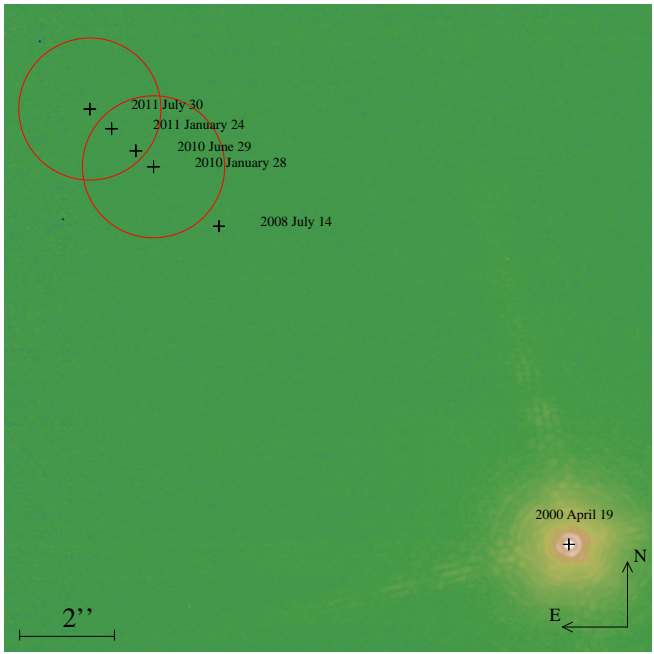}
\caption{\label{fig:cfht}
Canada France Hawaii Telescope (CFHT) K-band observation obtained 2000 April 19 using the adaptive optics bonnette (PUEO).  The ``+'' symbols indicate the position of the GJ 436 system for this observation and at each transit epoch of UCF-1.01.  Red circles indicate the minimum photometric aperture size (1.25 pixels) for which transit signals from the first and last confirmed events may still be clearly distinguished against the background noise.  We eliminate background objects within the overlapping region with $\Delta$K = 9.3 at a 5$\sigma$ confidence limit.
}
\end{figure}

GJ 436's large proper motion (across the sky) enables us to eliminate astrophysical false positives that could mimic the observed periodic decrease in flux.  Over our 1.5-year observational baseline of 4.5-{\micron} detections, the system moves $\sim$1.8$^{\prime\prime}$, equivalent to 1.5 pixels in {\em Spitzer}'s InfraRed Array Camera.  With aperture sizes as small as 1.25 pixels for the first (2010 January 28) and last (2011 July 30) observations, we find that the transit signals from UCF-1.01 are clearly distinguished against the background noise.  This limits the location of a potential background source \citep[such as an eclipsing-binary star system,][]{Torres2011} to the overlapping region within both apertures.  Using observations from the Very Large Telescope \citep[VLT, see Figure \ref{fig:vlt},][]{Rousset2003, Lenzen2003} and Canada France Hawaii Telescope \citep[CFHT, see Figure \ref{fig:cfht},][]{Rigaut1998} with adaptive optics imaging instruments at two different epochs, we eliminate background stars up to 12.7 and 9.3 magnitudes fainter, respectively, than GJ 436 at a confidence of 5$\sigma$.


\subsection{Instability Hypothesis}
\label{sec:althyp}

In this section we consider an alternate hypothesis to that of detecting two sub-Earth-sized exoplanets, that the observed variations are the result of instrumental or stellar instabilities.  We begin by calculating the probability of observing UCF-1.01 and UCF-1.02 by chance.  Then, we quantify the occurrence rate of transit-like instabilities and estimate the probability that these instabilities are periodic.  Finally, we compare our model fits to the null hypothesis, which is expressed by a model that does not contain planet parameters, to see if the additional free parameters are justified.

In search of transit signals in the GJ 436 system, we examined 11 light curves at 3.6 and 4.5 {\microns} (not counting the 2011 July 30 dataset in which we predicted the transit).  Both channels have the photometric stability necessary to detect GJ 436c.  Of the 71.3 hours of data, there are eight transit or eclipse events of GJ 436b, each lasting $\sim$1-hour in duration.  Since we cannot distinguish overlapping transits, we have 63.3 hours of usable data with an average light-curve duration of $\sim$5.75 hours.  Given that the period of UCF-1.01 is 1.3659 days, the probability that a transit will occur in a typical event is 17.5\%.  Using the binomial distribution, we calculate a 30.1\% chance of observing three or more transits of UCF-1.01 in the 11 available light curves.  Recall that our fourth transit event of UCF-1.01 was predicted, rather than occurring by chance, so it does not enter into the calculation.

We repeat the above calculation for UCF-1.02 but must first estimate its orbital period by considering the transit duration ratio between itself and GJ 436b.  We find that the durations are nearly identical; however, both planets are unlikely to occupy the same orbit.  So, we perform two sets of calculations: one for each side of the 1-sigma uncertainty in UCF-1.02's transit duration.  Using a period of 5.563 days, the probability of observing two or more UCF-1.02 transits is 7.9\%.  Using a period of 1.785 days, the probability of observing two or more transits increases to 44.6\%.  The combined probability of observing both planets ranges from 2.3\% to 13.4\%.

We compare these results to the alternate hypothesis, namely that the observed variations are the result of instrumental or stellar instabilities.
To begin, we analyzed nearly 120 hours of GJ 1214 data at 4.5 {\microns} ({\em Spitzer} program 70049).  This M dwarf is similar to GJ 436 and should exhibit similar levels of activity (stellar instabilities).  
If the instabilities are instrumental, then it should not matter which star we analyze unless the instabilities are flux-dependent (GJ 436 is almost 3 magnitudes brighter than GJ 1214 in the infrared).
From our GJ 1214 light-curve results, we identify two transit-like events based on their depths ($>$200 ppm) and durations ($\sim$1 hour).  Assuming these events are not the result of planet transits, for any given hour of 4.5-{\micron} observations, there is a 1.7\% probability of having an instability event.  
We then apply the binomial distribution to determine that the probability of detecting five or more instabilities in 63.3 hours of data is 0.42\%.  Recall that we do not count the 2011 July 30 dataset or use times during GJ 436b transits/eclipses.  If we assume that the instabilities only appear at 4.5 microns, the probability of detecting five or more instabilities in 44.7 hours of data decreases to 0.088\%.  

Since we cannot find a physical mechanism for reducing the stellar flux in a transit-like way with a repeatable period, any observed instabilities must be random events.  We consider the probability of detecting four random instability events that happen to coincide with a given period (in this case, 1.3659 days).  The first two instability events establish the ``period'' under consideration.  As calculated above, the third and fourth instability events each have a 1.7\% probability of occurring within 30 minutes of the established period (total time window is 1 hour).  Their combined probability is 0.029\%.

We conclude that the single-planet hypothesis is 72 times more likely than the most favorable instrumental/stellar-instability scenario and 1040 times more likely than detecting four random instability events that happen to coincide with a given period.  The two-planet hypothesis is 5.5 -- 32 times more likely than the most favorable instrumental/stellar-instability scenario and 79 -- 460 times more likely than detecting four random instability events that happen to coincide with a given period.

Finally, we test the strength of our two sub-Earth-sized exoplanet candidates by comparing various fits to the null hypothesis, which asserts that there are no new planets.  Recall that a lower $\Delta$BIC value indicates that the additional free parameters are warranted in the model fit.
Relative to the null hypothesis, $\Delta$BIC decreases by 11.4 when we include UCF-1.01's transit parameters in a joint model fit. Alternatively, if we add only UCF-1.02's transit parameters then $\Delta$BIC increases by 36.2.  Including both planets' transit parameters in a joint model fit results in an increase in $\Delta$BIC of 14.5, relative to the null hypothesis.
Thus, the BIC favors a model that includes UCF-1.01 but disfavors models that include UCF-1.02.  This result is directly related to the number of observed transits for each planet candidate and indicates that we need to obtain more than two transit observations of UCF-1.02 to increase the detection significance and surpass the BIC's penalty for additional free parameters.  We conclude that the available data support UCF-1.01 as a strong exoplanet candidate and signify that UCF-1.02 is a weak exoplanet candidate.

\subsection{Radial-Velocity Constraints}

The 3.6-meter ESO telescope at La Silla Observatory \citep{Mayor2003,Pepe2004} utilized the HARPS spectrograph with the settings described by \citet{Bonfils2011} to obtain 171 observations of GJ 436 at 550 nm.  Xavier Bonfils (personal communication) provided us with the extracted, unpublished RV measurements so that we could attempt to constrain the mass of UCF-1.01.  We retained 159 data points (12 were removed due to the Rossiter-McLaughlin effect).  To these data, we added 
41 GJ 436b primary transit times \citep[and references therein]{Knutson2011}, 14 GJ 436b secondary eclipse times \citep{Stevenson2010, Knutson2011}, and an 8.0-{\micron} photometric light curve from \citet{Deming2007}.  The light curve (retrieved from the Infrared Processing and Analysis Center, IPAC) is binned into 445 points and normalized to remove the time-dependent ramp.

We apply a two-planet model with the transit ephemeris for the second planet fixed to the best-fit value listed in Table \ref{tab:aei} and the eccentricity fixed to 0.  The fit utilizes the empirical stellar density calibration of \citet{Enoch2010} to determine the stellar mass, in addition to other system parameters, granting this fit a much broader scope than the modeling described by \citet{Campo2011}.  
We employ a Levenberg-Marquardt minimizer to find the best-fit parameters to our model and a Markov-chain Monte Carlo routine with 10\sp{6} iterations to estimate uncertainties.  We express our $\chi^2$ function as follows:

\begin{equation}
\begin{split}
  \chi^2 =  \sum_{i} \left [\frac{v_{i}-\overline{v_{i}}}{\sigma_{v,i}} \right ]^2 +
            \sum_{j} \left [\frac{t_{j}-\overline{t_{j}}}{\sigma_{t,j}} \right ]^2  \\
         +\;\sum_{k} \left [\frac{o_{k}-\overline{o_{k}}}{\sigma_{o,k}} \right ]^2 +
            \sum_{m} \left [\frac{p_{m}-\overline{p_{m}}}{\sigma_{p,m}} \right ]^2
\end{split}
\end{equation}

\noindent where $v$, $t$, $o$, and $p$ represent the HARPS radial velocities, primary-transit times, secondary-eclipse times, and photometric data, respectively.  The over-lined quantities indicate computed values and $\sigma$ represents the uncertainty for each measurement.
We adjust for transit-eclipse light travel times and for leap seconds in this fit.
Using the above data with an estimated stellar jitter of 1.7 m/s, we do not detect the signal of the second planet but cannot repudiate its existence.  The 3$\sigma$ upper limit of the semi-amplitude is 0.6 m/s, corresponding to an upper limit of 0.6 $M\sb{\oplus}$, which is larger than our mass constraints using the density arguments below.

\subsection{Mass Constraints}

\begin{figure}
\clearpage
\begin{center}
\includegraphics[width=\columnwidth, clip]{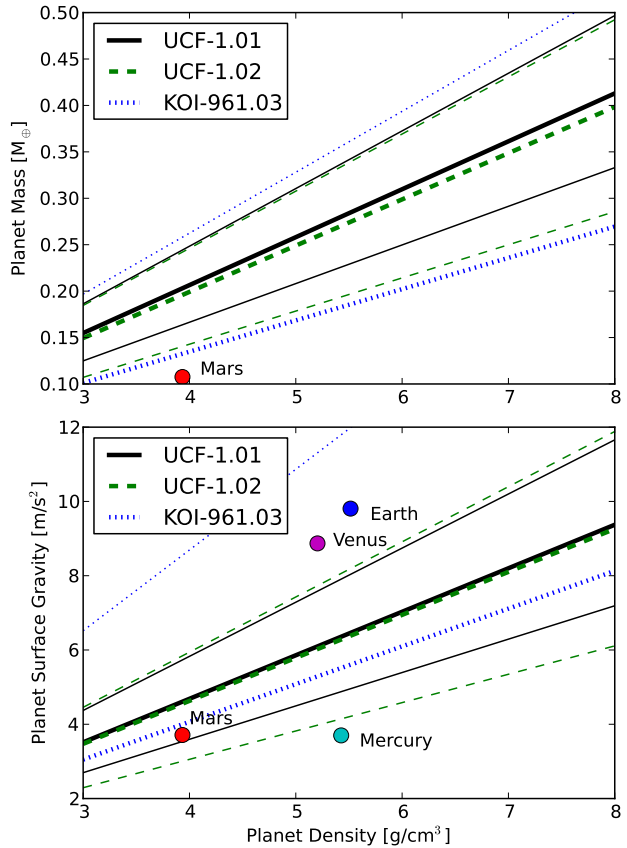}
\end{center}
\caption{\label{fig:mass}
Mass and surface-gravity constraints on UCF-1.01 (solid lines) and UCF-1.02 (dashed lines).  Bold lines depict the best-fit values and thin lines depict the upper and lower 1$\sigma$ uncertainties.  Exoplanet KOI-961.03 and solar-system planets are included for reference.
}
\end{figure}

%
Unable to effectively constrain the mass of UCF-1.01 using RV data, we consider a range of possible bulk densities for a terrestrial-sized planet (see Figure \ref{fig:mass}).
Given a mean bulk density between 3 and 8 g cm$^{-3}$, we limit the mass of UCF-1.01 to be 0.15 -- 0.40 $M\sb{\oplus}$ and estimate the surface gravity to be 0.36 -- 0.94 $g$.  We place similar limits on the mass and surface gravity of UCF-1.02.  Assuming an Earth-like density of 5.515 g cm$^{-3}$, we estimate masses of 0.28 and 0.27 $M\sb{\oplus}$ for UCF-1.01 and UCF-1.02, respectively, which correspond to surface gravities of $\sim$0.65 times that on Earth.

\subsection{Orbital Constraints}
\label{sec:orbit}

UCF-1.01 may exhibit TTVs due to gravitational interactions with GJ 436b in a near-2:1 orbital resonance or with UCF-1.02, which has an unknown orbit.  This may explain why UCF-1.01's transit time is 20 minutes early in the 2010 January 28 data set; however, the parameter's probability distribution is bimodal (see Figure \ref{fig:cp21corr}) and the other peak is only 6 {\pm} 7 minutes early.  The three remaining UCF-1.01 transit times occur within five minutes of their predicted times and have a typical uncertainty of {\pm}3 minutes.  More precise observations could establish whether these deviations are TTVs.  


Using the known orbital parameters of GJ 436b and UCF-1.01, we performed orbital-stability simulations using the Mercury numerical integrator \citep{Chambers1999}.  
Assuming an Earth-like density of 5.5 g cm\sp{-3}, the predicted mass of UCF-1.01 is 0.28 M\sb{$\oplus$}.  We supplied the code with initial starting conditions, listed in Table \ref{tab:merc}, based on transit times from the 2011 January 24 dataset.  Our results indicate that the orbits are stable out to at least 100 Myr.  The best-fit line shows a change in semi-major axis of 5.3e-06 au/Gyr.
The osculating UCF-1.01 orbital parameters exhibit a periodic trend every $\sim$35 years wherein the eccentricity varies between 0 and 0.21, the peak-to-trough inclination amplitude is 3.2$^{\circ}$, and TTVs vary from {\pm}200 to {\pm}3 minutes.
A $\sim$40-day periodic trend is also evident but with smaller variations in the osculating orbital parameters.  Due to UCF-1.01's relatively small mass, variations in GJ 436b's orbital parameters over the 35-year timespan are below {\em Spitzer}'s sensitivity limits.  Next-generation facilities may be able to constrain UCF-1.01's orbital parameters through improved RV measurements or by measuring its time of secondary eclipse.

\begin{table}[ht!]
\caption{\label{tab:merc} 
Initial orbital parameters for GJ 436b and UCF-1.01}
\centering
\begin{tabular}{ccc}
    Parameter                           &   GJ 436b         &   UCF-1.01         \\
    \hline
    Semi-major Axis                     &   0.0287 au       &   0.0185 au       \\
    Eccentricity                        &   0.1371          &   0               \\
    Inclination                         &   86.699$^{\circ}$&   85.1$^{\circ}$  \\
    Argument of Periapsis               &   351.0$^{\circ}$ &   0.0$^{\circ}$   \\
    Longitude of the Ascending Node     &   0.0$^{\circ}$   &   0.0$^{\circ}$   \\
    Mean Anomaly                        &   282.6$^{\circ}$ &   90$^{\circ}$    \\
    \hline
\end{tabular}
\end{table}

\subsection{Atmospheric Constraints}
\label{sec:atm}

UCF-1.01 is unlikely to have retained any original atmosphere due to its weak gravitational field, close proximity to its host star, and estimated 6-Gyr age of the system \citep{Torres2007}. The planet receives a substantial soft x-ray and extreme ultraviolet (XUV) flux; we estimate 700 -- 900 erg cm$^{-2}$ s$^{-1}$ \citep{Sanz-Forcada2011, Ehrenreich2011}, or $\sim$1,000 times the present XUV flux received by the Earth. 
Such an intense XUV flux leads to a very hot thermosphere and subsequent hydrodynamic escape \citep{Tian2009}.
Shortly after formation, outgassing from an Earth-like, silicate-rich mantle could have produced an initial water-vapor-rich atmosphere for UCF-1.01 \citep{Schaefer2011}.  However, the water vapor would readily have been photolyzed by ultraviolet radiation at high altitudes, leading to a hydrogen-dominated thermosphere that likely extended to the planet's Roche distance of $\sim$25,000 km \citep{Erkaev2007}, given the planet's low gravity.  
In this situation, the mass-loss rate for energy-limited hydrodynamic flow \citep{Erkaev2007} implies a hydrogen loss rate of about $8 \times 10^{10}$ g s\sp{-1} (assuming an XUV heating efficiency of 1), or 1.4 times the planet's mass lost in 1 Gyr.  This indicates that hydrogen was lost from UCF-1.01's atmosphere very early in its history.  Some heavy elements would have been entrained in the hydrodynamic flow, but the early atmosphere would have become increasingly oxidized as hydrogen was lost. Carbon dioxide could then have dominated at some later point in the atmosphere's history, but even a CO$_2$-rich atmosphere would be unstable. Scaling from hydrodynamic models \citep{Tian2009}, we estimate that carbon would be lost from a CO$_2$-rich atmosphere at $\sim1 \times 10^8$ g s\sp{-1}, or 1\% of the planet's mass over its lifetime -- an amount likely greater than the planet's initial inventory of CO$_2$.  Atmospheres dominated by molecular nitrogen or oxygen would be lost on even shorter timescales \citep{Tian2009}.

UCF-1.01 could support a transient, present-day atmosphere if recent impacts were to deliver volatiles rather than preferentially erode any atmosphere, or if tidal heating were to supply volatiles from the crust/mantle. The latter scenario is particularly attractive if a recycling mechanism exists for any heavy atmospheric constituents (e.g., volcanic emission of sulfur dioxide, followed by photolysis to sulfur and oxygen atoms, dayside-to-nightside transport, condensation, and subsequent melting and re-vaporization of sulfur deposits).  In this speculative scenario, UCF-1.01 could resemble a hot Io that has lost its lighter and more volatile elements.  Any transient atmosphere will likely have a low surface pressure and be highly extended, which could fill the Roche lobe and/or produce a tail.
Transit observations at ultraviolet wavelengths could confirm or rule out such an extended atmosphere, and one might search particularly at wavelengths in which atomic and ionized sulfur and oxygen would be expected to absorb. Given that volcanically supplied sodium and potassium might be transient atmospheric constituents, visible-wavelength transit observations might also prove useful.

\section{CONCLUSIONS}
\label{sec:concl}

In this paper, we announced the detection of UCF-1.01 and UCF-1.02, two sub-Earth-sized transiting exoplanet candidates orbiting the nearby M dwarf GJ 436.  Their detections were possible with BLISS mapping and Time-series Image Denoising (TIDe), the latter of which is a novel wavelet-based technique that decreases high-frequency noise in short-cadence, time-series images to improve image centering precision.
We presented four transits of UCF-1.01 and two transits of UCF-1.02 at 4.5 {\microns}, an independent analysis that confirms our best-fit results within 1.5$\sigma$, an 8.0-{\micron} phase curve of GJ 436b that includes transits of UCF-1.01, and EPOXI data that are consistent with the presence of a sub-Earth-sized exoplanet.
To definitively establish UCF-1.01 as a planet (to be called GJ 436c), we require only a few hours of additional observations, preferably from another telescope or at least at a different wavelength.  Establishing UCF-1.02 as a planet (to be called GJ 436d) would likely require an extended observing campaign to constrain its period then successfully predict a transit.
Finally, we confirmed the GJ 436b 4.5-{\micron} results presented by \citet{Stevenson2010} through an additional non-detection during secondary eclipse; however, we were unable to confirm the strong eclipse depth at 3.6 {\microns} due to stellar activity.  The current data still support a methane-deficient and carbon monoxide-rich dayside atmosphere.

\acknowledgments

We thank Tom Loredo for providing comments on TIDe and members of the HARPS team for discussions and for providing unpublished RV data with an independent analysis.
 We acknowledge Drake Deming for supplying the GJ 1214 data and leading the EPOXI mission; we also acknowledge Heather Knutson, the PI of {\em Spitzer} program 50056.
 We thank contributors to SciPy, Matplotlib, and the Python Programming Language, the free and open-source community, the NASA Astrophysics Data System, and the JPL Solar System Dynamics group for software and services. This work is based on observations made with the {\em Spitzer Space Telescope}, which is operated by the Jet Propulsion Laboratory, California Institute of Technology under a contract with NASA. Support for this work was provided by NASA through an award issued by JPL/Caltech.
\\

\bibliography{ms}

\newpage
\begin{appendices}

\section{Correlation Plots and Histograms}
\label{sec:corr}

\begin{figure}[h]
\includegraphics[clip,width=0.5\textwidth]{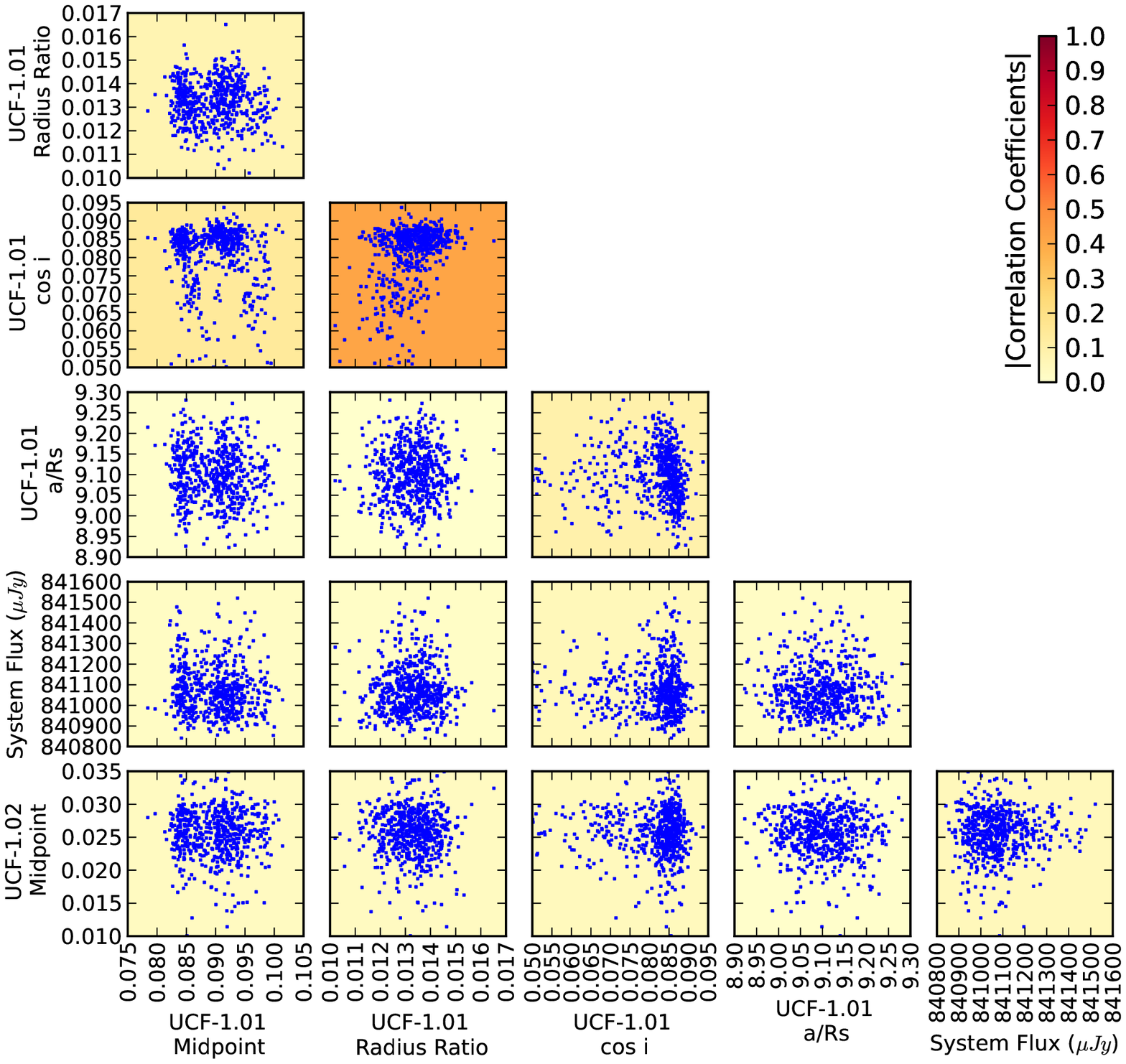}\hfill
\includegraphics[clip,width=0.5\textwidth]{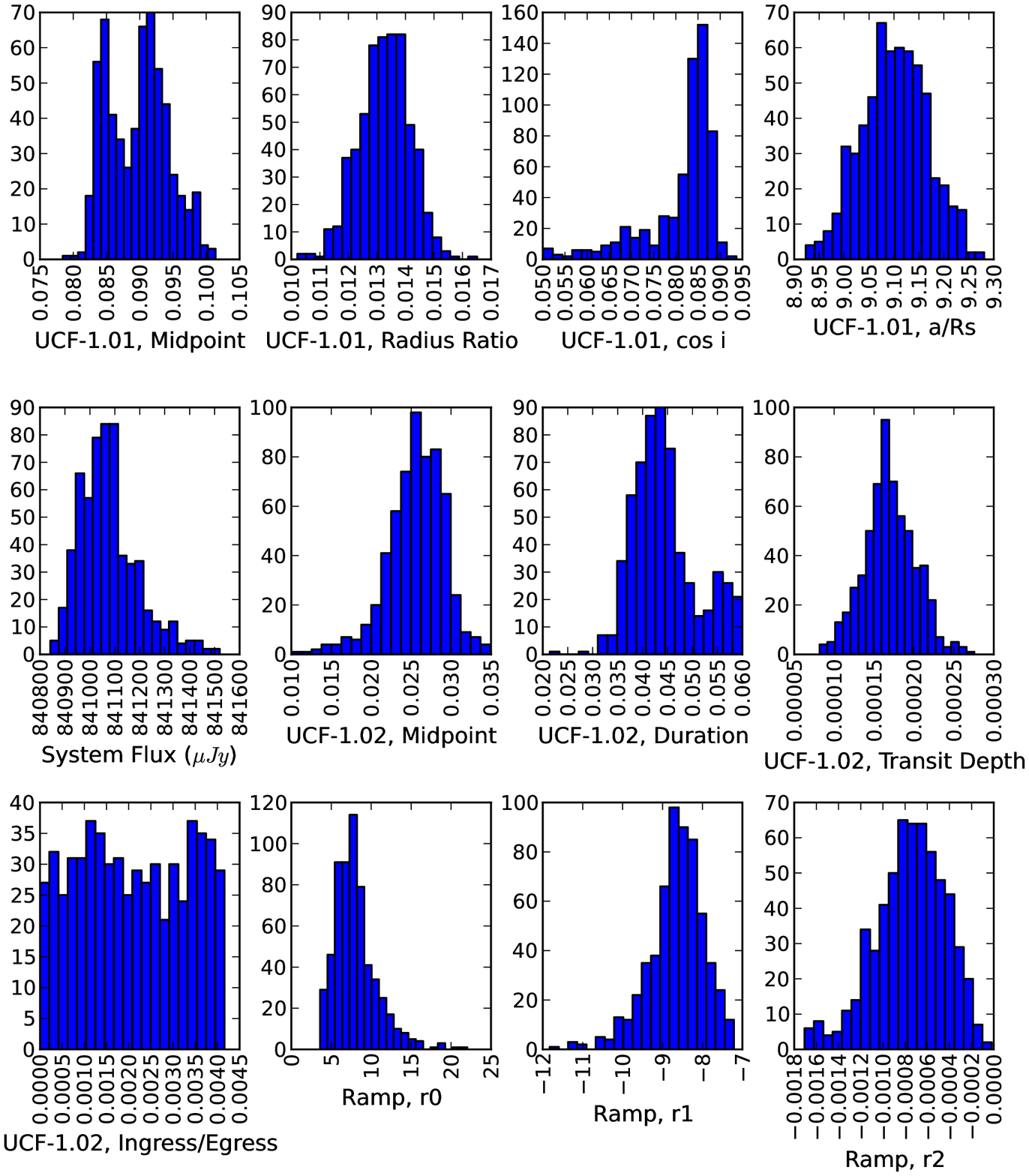}\hfill
\includegraphics[clip,width=0.5\textwidth]{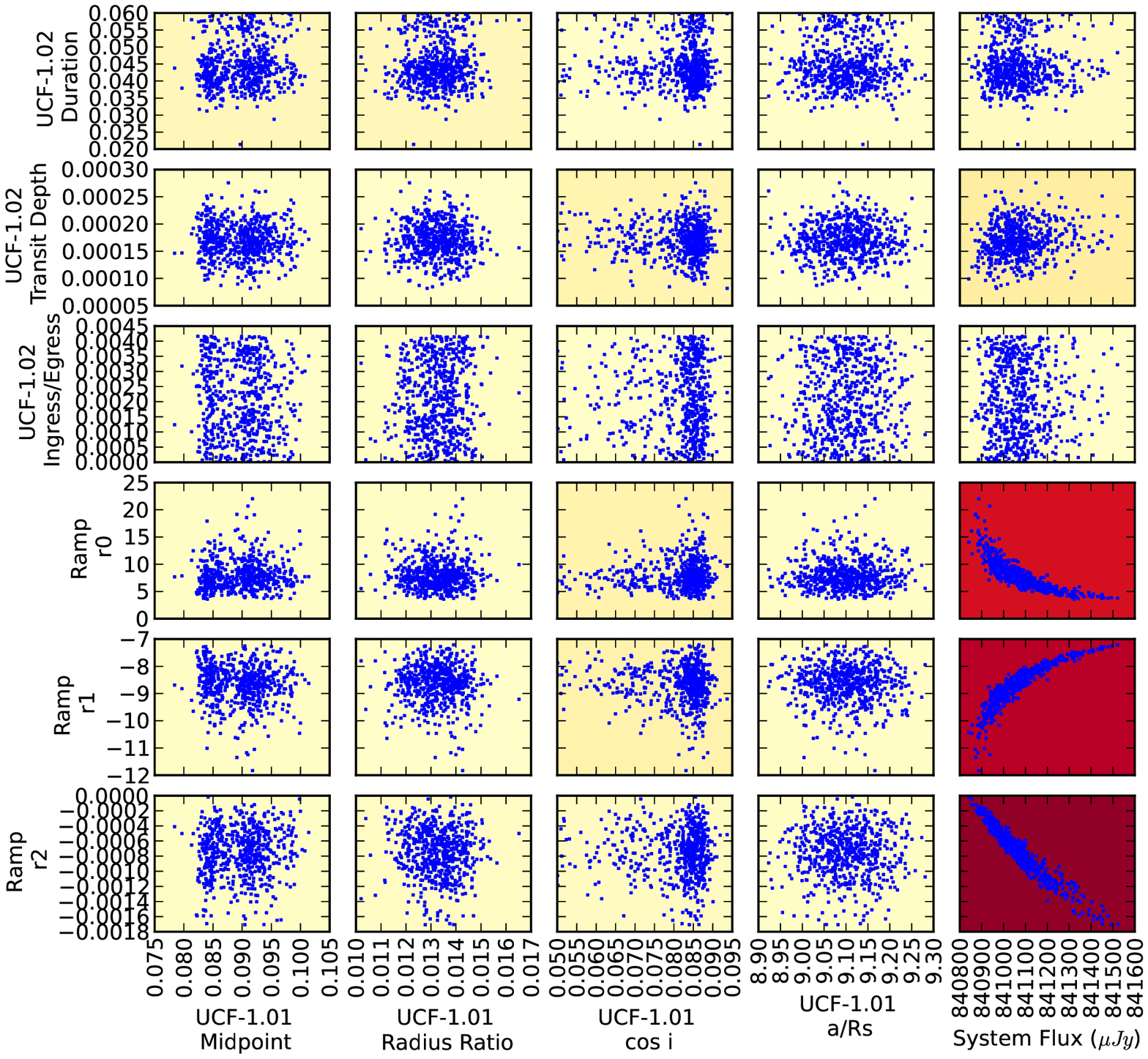}\hfill
\includegraphics[clip,width=0.5\textwidth]{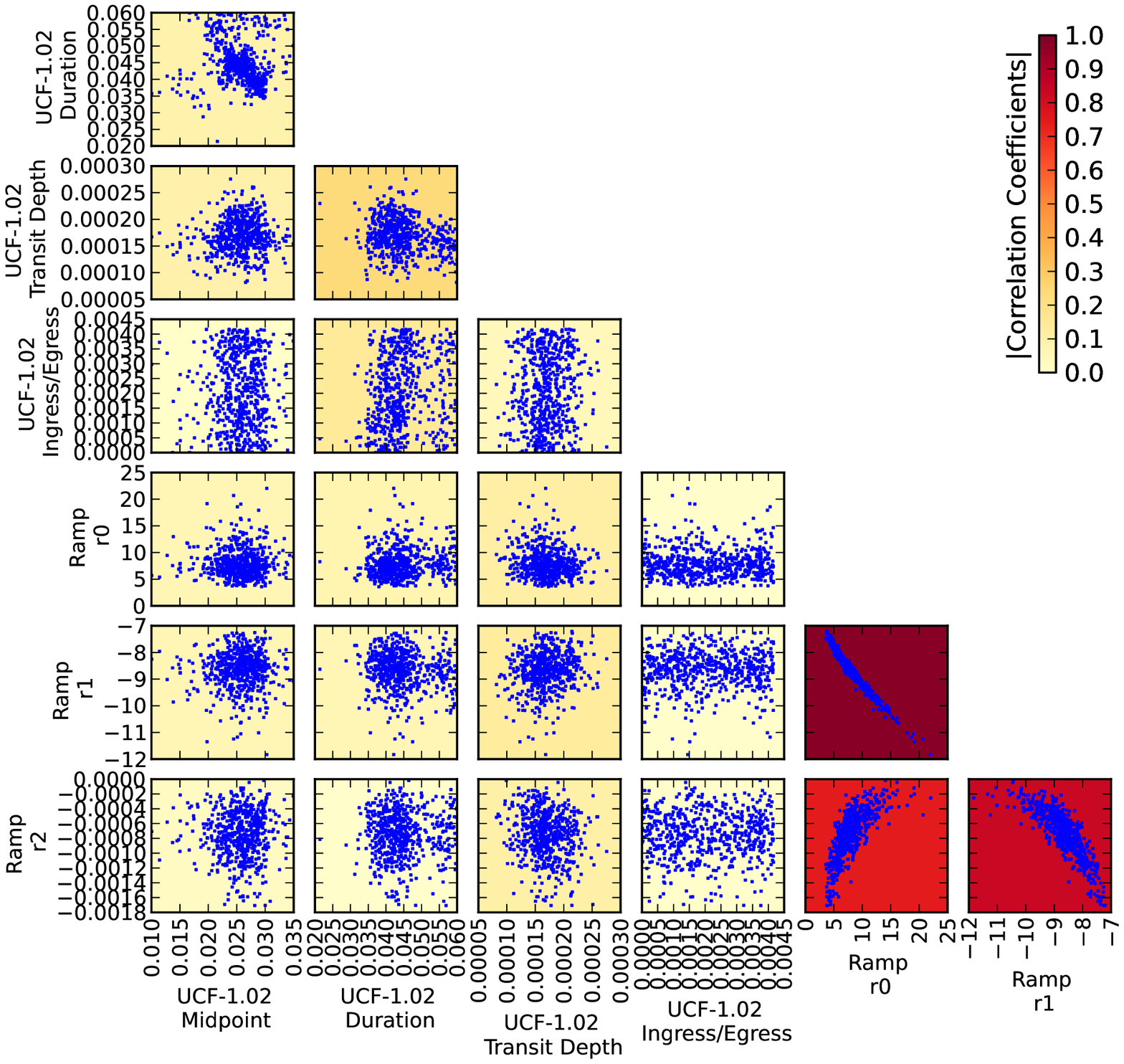}\hfill
\caption{\label{fig:cp21corr}{\small
Correlation plots and histograms for the 18-hour 2010 January 28 {\em Spitzer} observation containing transits of UCF-1.01 and UCF-1.02.  We plot every 4000\sp{th} step in the MCMC chain to decorrelate parameter values.  UCF-1.01's distribution of mid-transit times (midpoints) is bimodal, so we favor the median value over the best-fit value (see Table \ref{tab:aei}).  UCF-1.02's ingress/egress times are unconstrained from our model fit.
}}
\end{figure}

\clearpage
\begin{figure}[h]
\includegraphics[clip,width=0.5\textwidth]{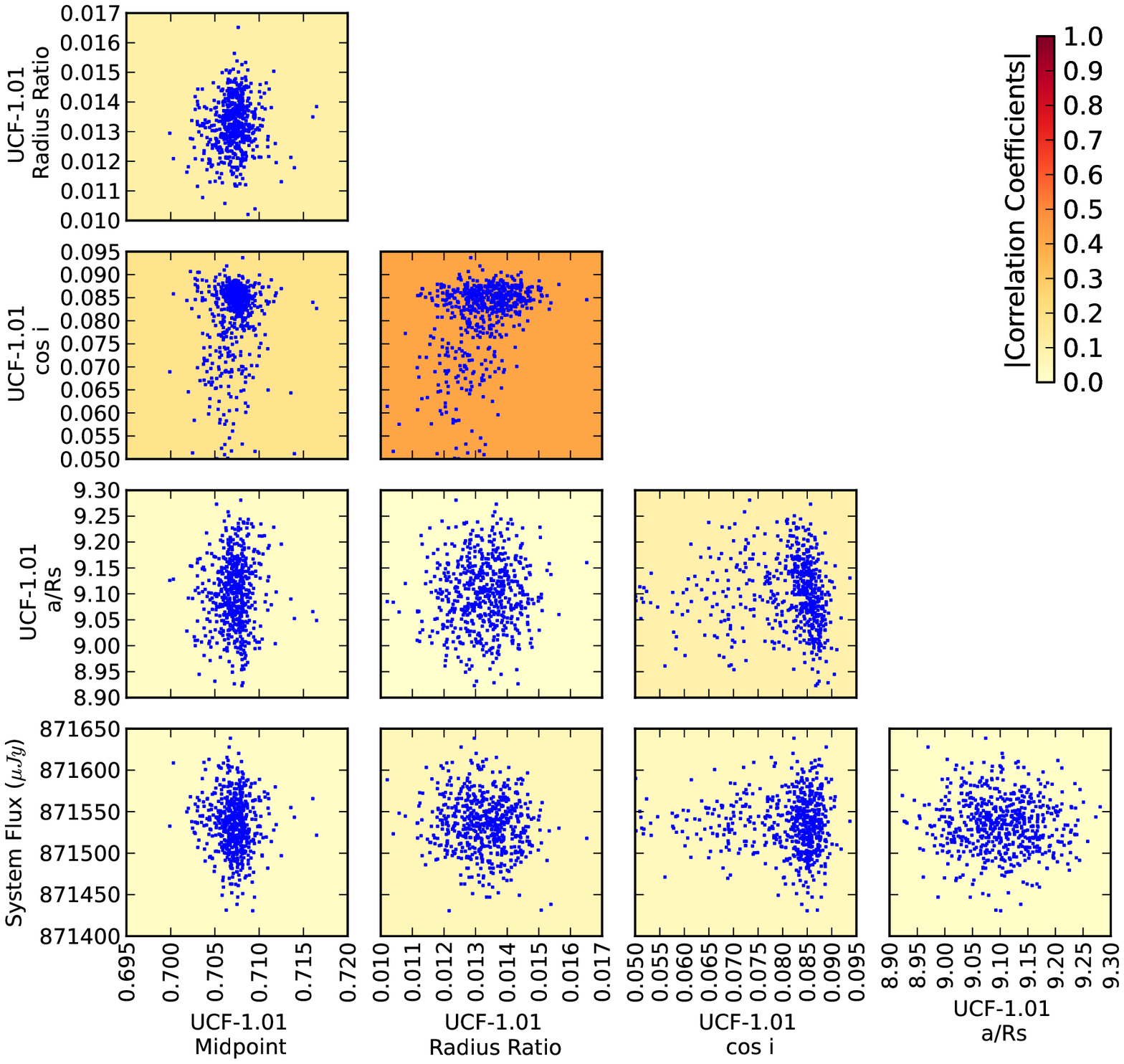}\hfill
\includegraphics[clip,width=0.5\textwidth]{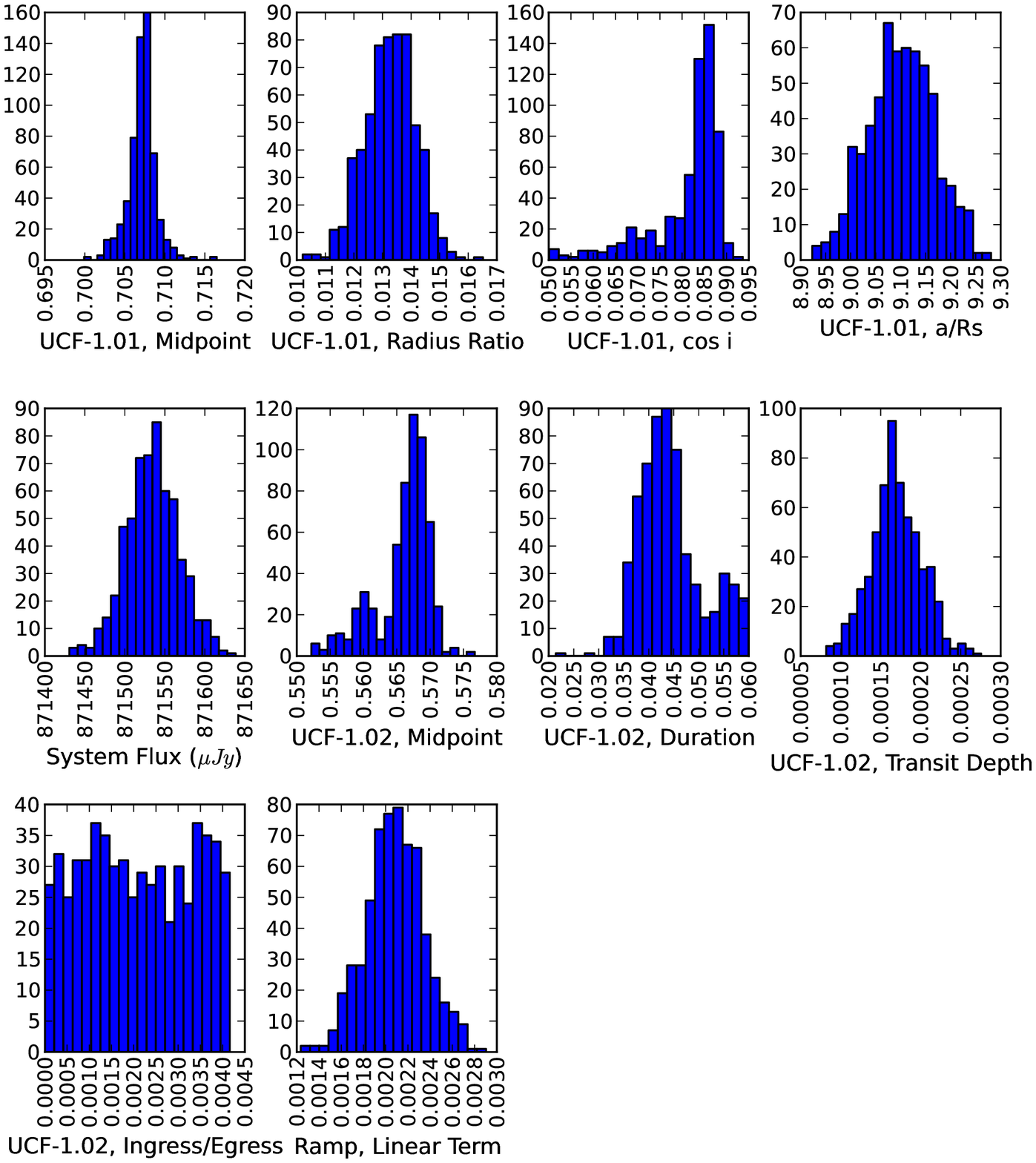}\hfill
\includegraphics[clip,width=0.5\textwidth]{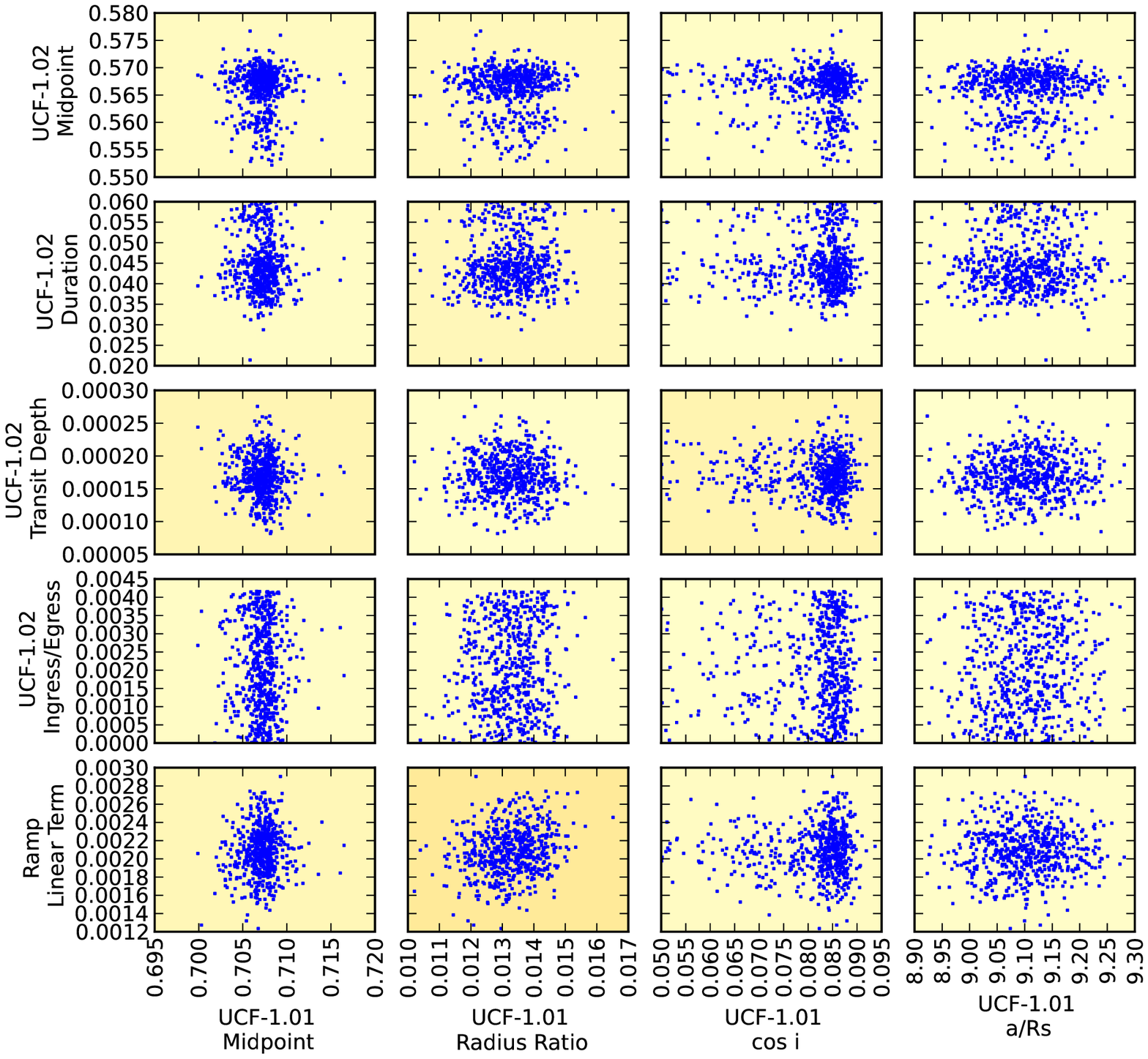}\hfill
\includegraphics[clip,width=0.5\textwidth]{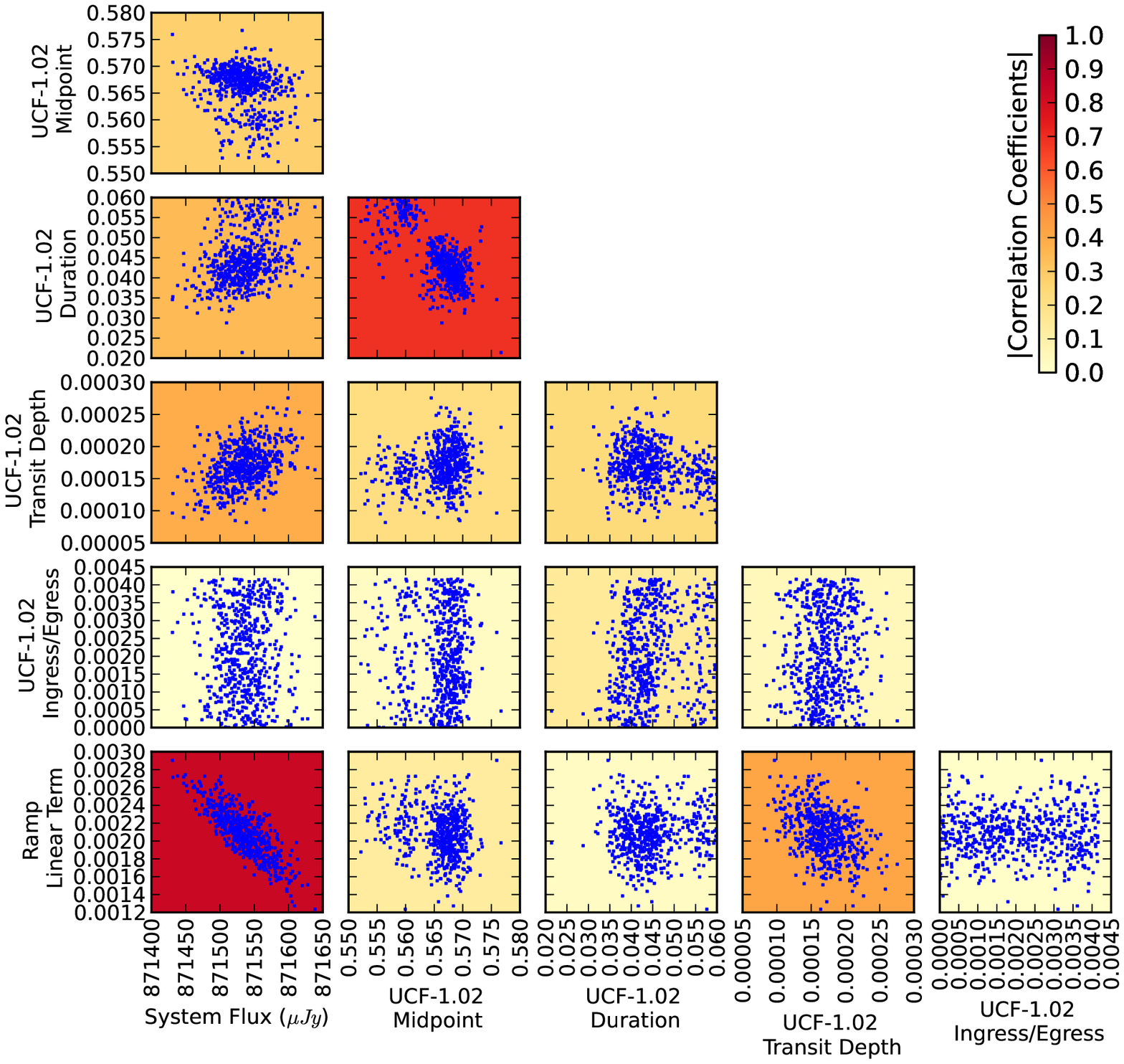}\hfill
\caption{\label{fig:bs22corr}{\small
Correlation plots and histograms for the 2010 June 29 {\em Spitzer} observation containing transits of UCF-1.01 and UCF-1.02.  We plot every 4000\sp{th} step in the MCMC chain to decorrelate parameter values.  UCF-1.02's ingress/egress times are unconstrained from our model fit.
}}
\end{figure}

\clearpage
\begin{figure}[h]
\includegraphics[clip,width=0.5\textwidth]{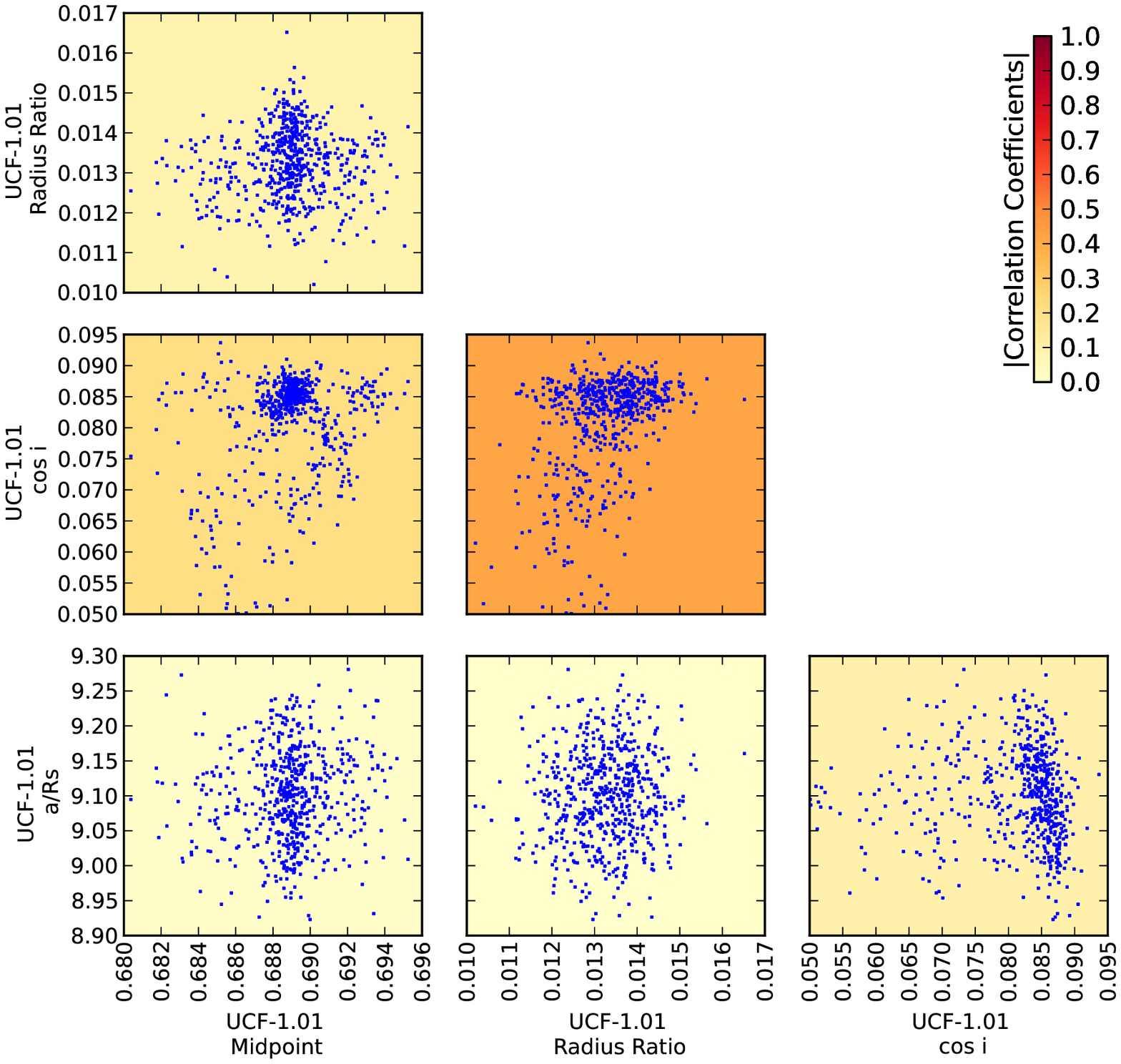}\hfill
\includegraphics[clip,width=0.5\textwidth]{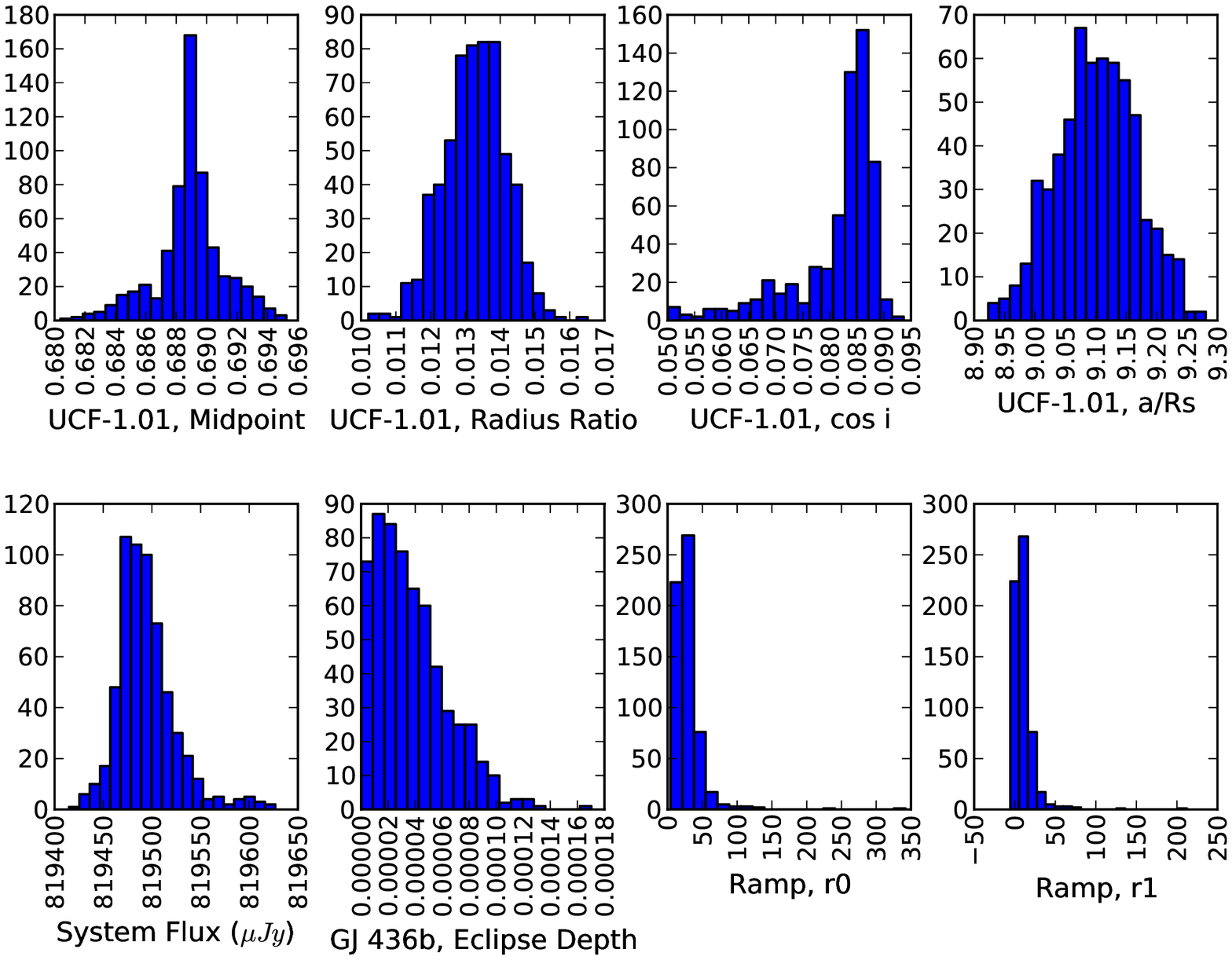} \hfill
\includegraphics[clip,width=0.5\textwidth]{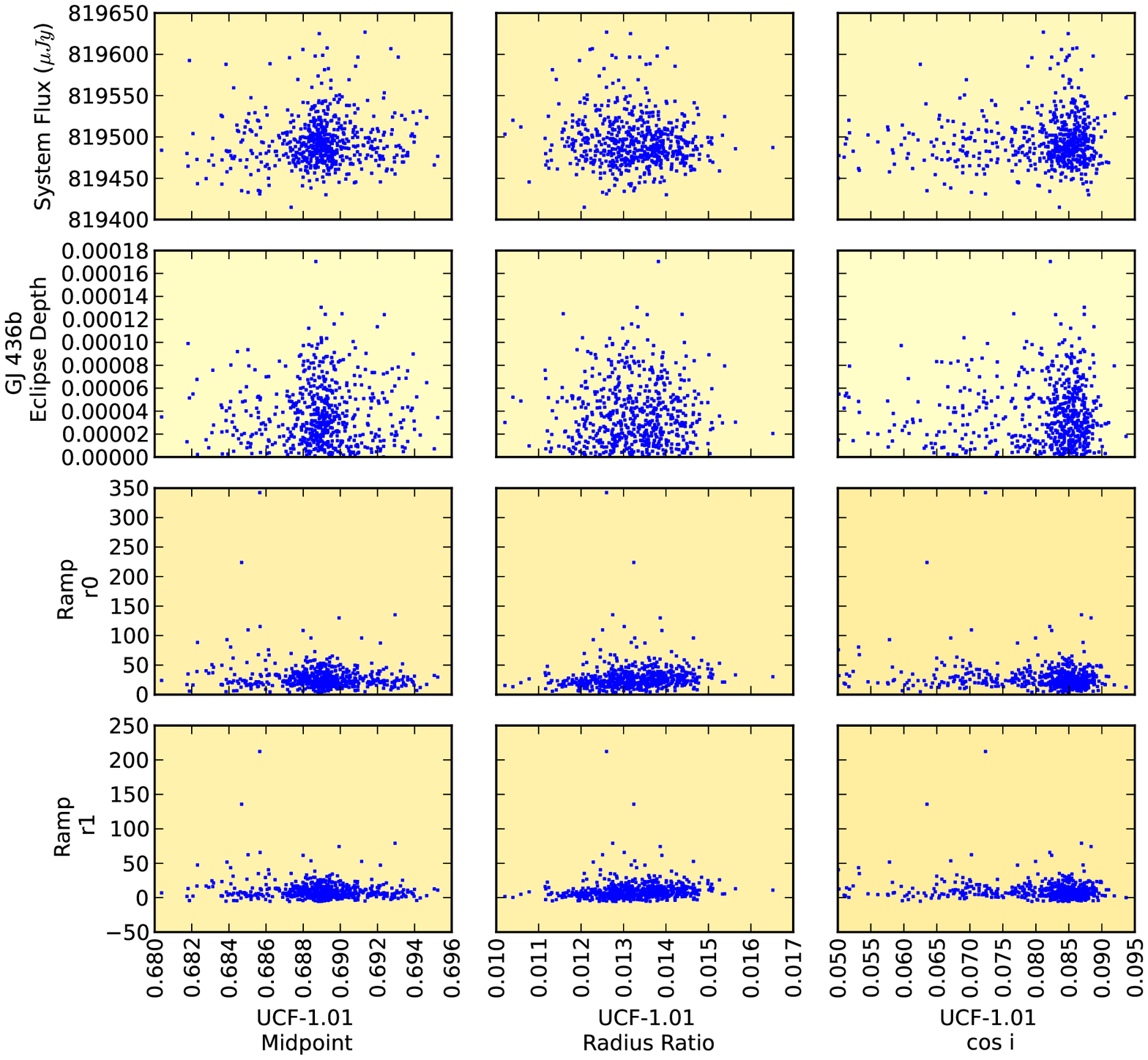}\hfill
\includegraphics[clip,width=0.5\textwidth]{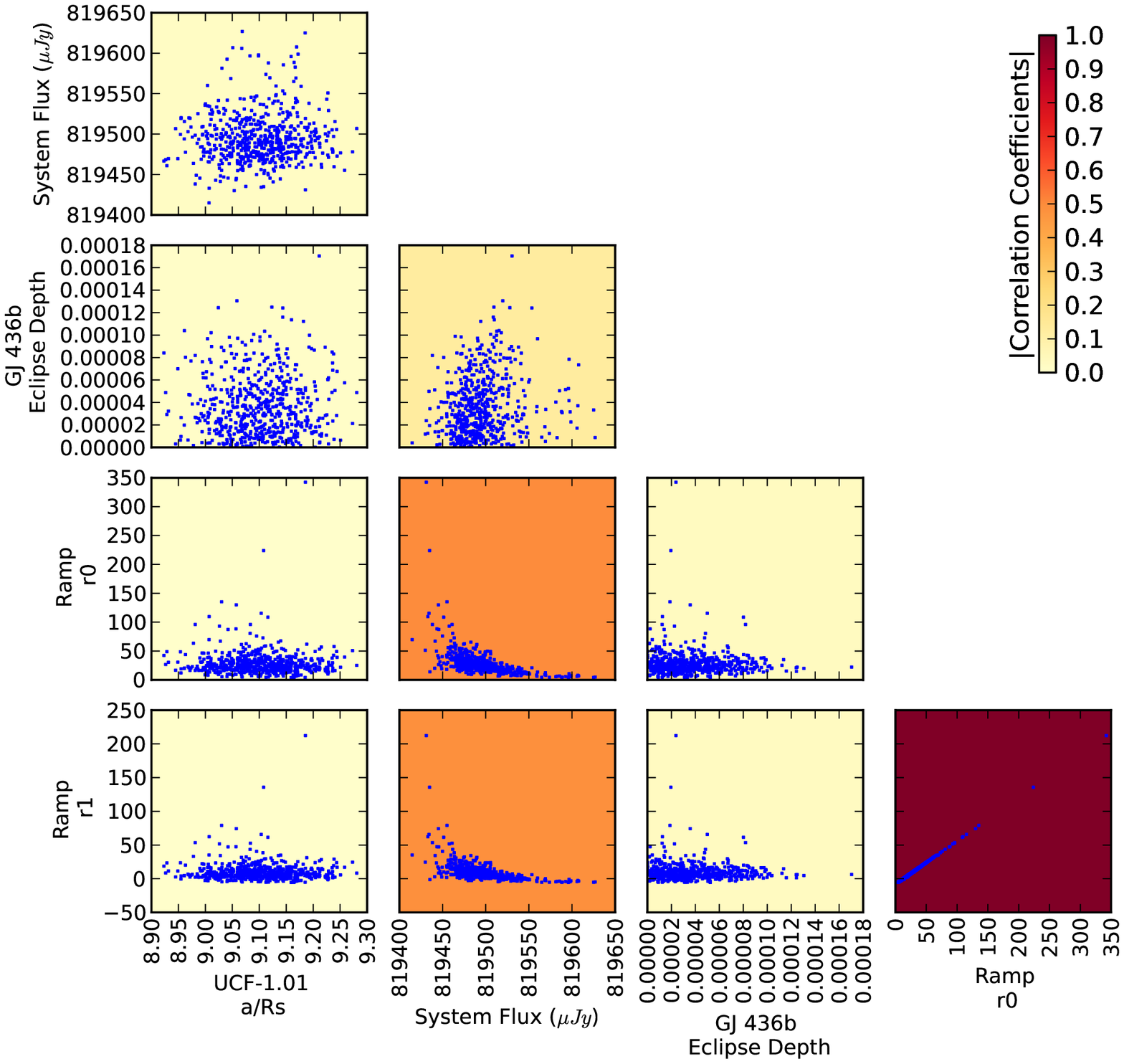}\hfill
\caption{\label{fig:bs23corr}{\small
Correlation plots and histograms for the 2011 January 24 {\em Spitzer} observation containing a transit of UCF-1.01 and an eclipse of GJ 436b.  We plot every 4000\sp{th} step in the MCMC chain to decorrelate parameter values.
}}
\end{figure}

\clearpage
\begin{figure}[h]
\includegraphics[clip,width=0.5\textwidth]{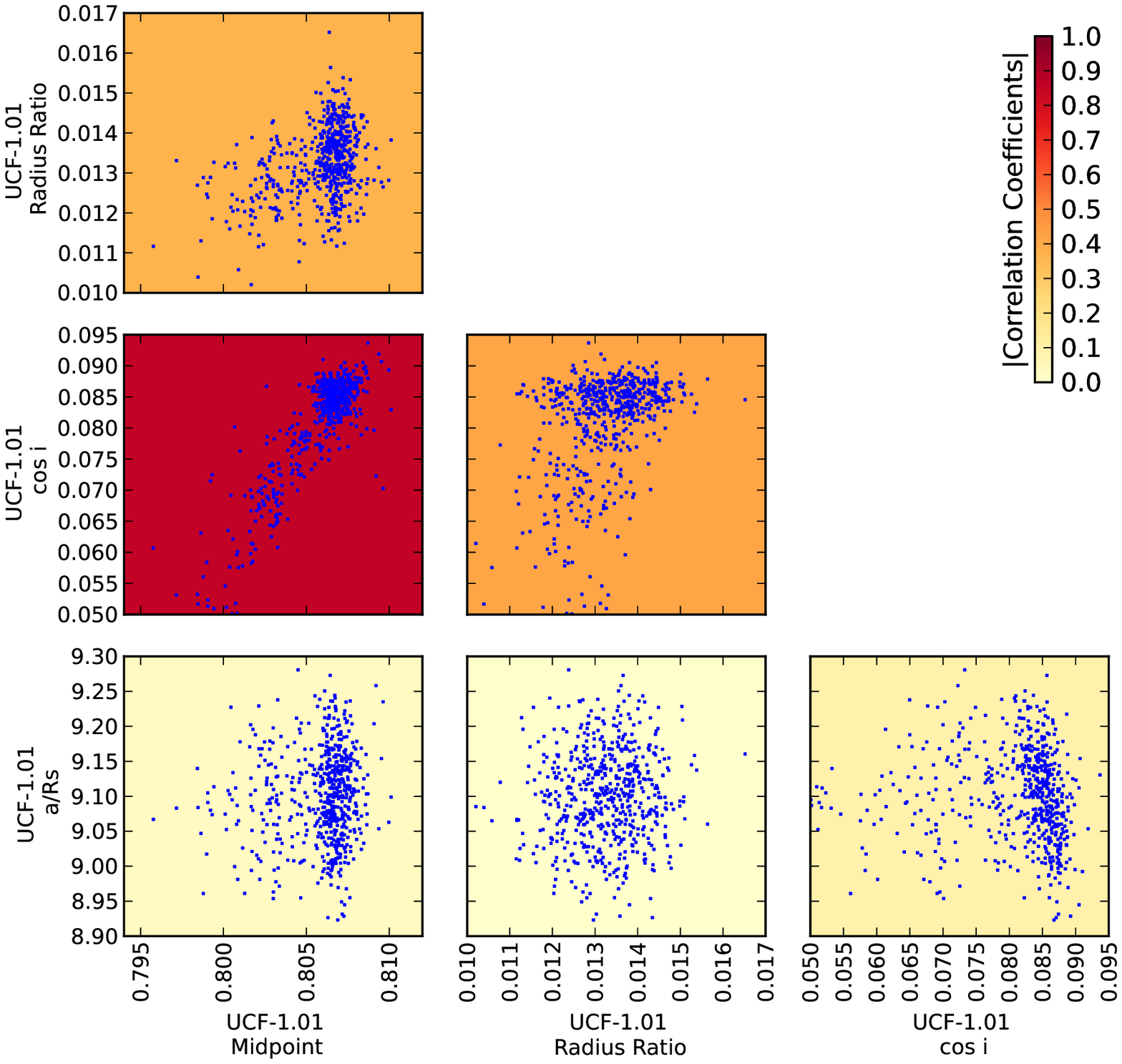}\hfill
\includegraphics[clip,width=0.5\textwidth]{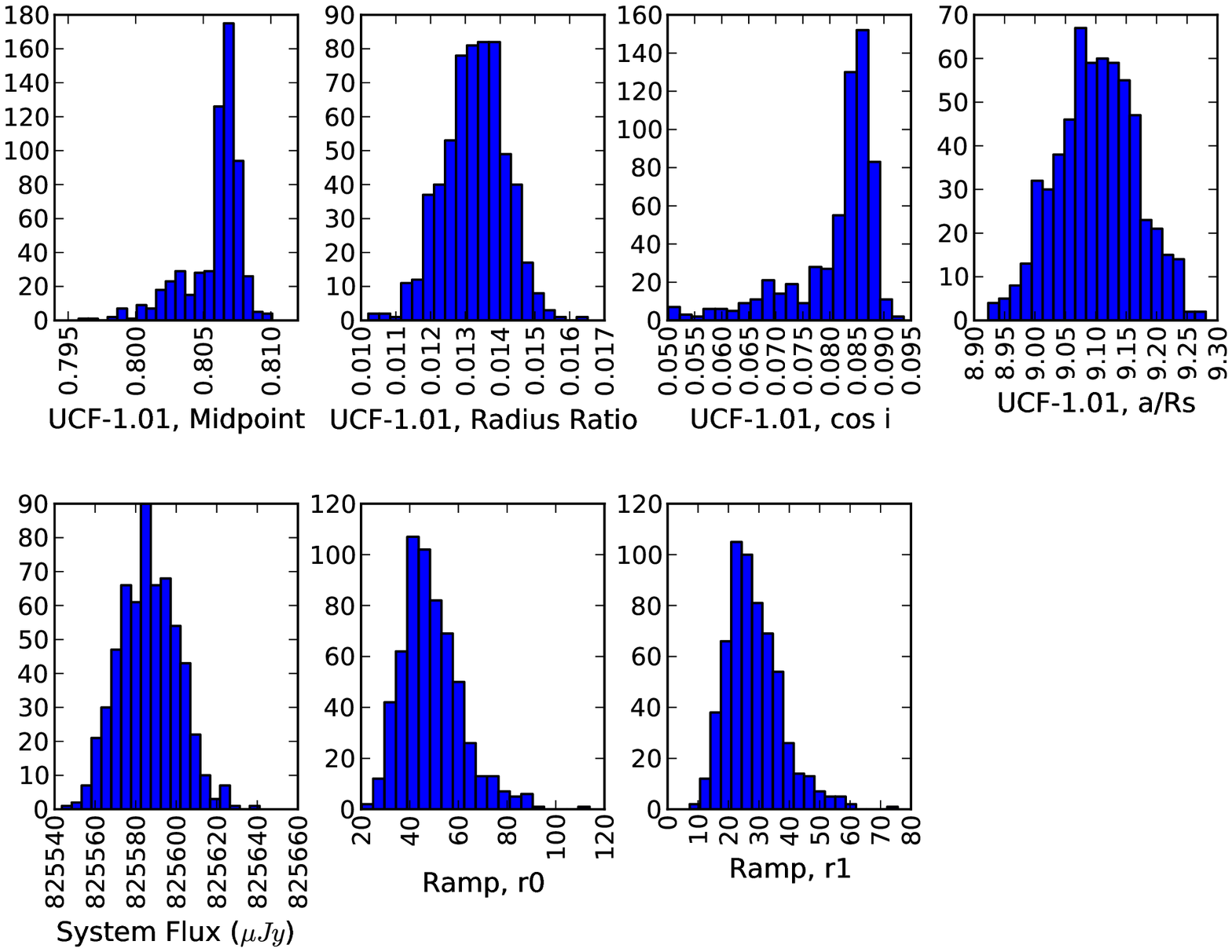}\hfill
\includegraphics[clip,width=0.5\textwidth]{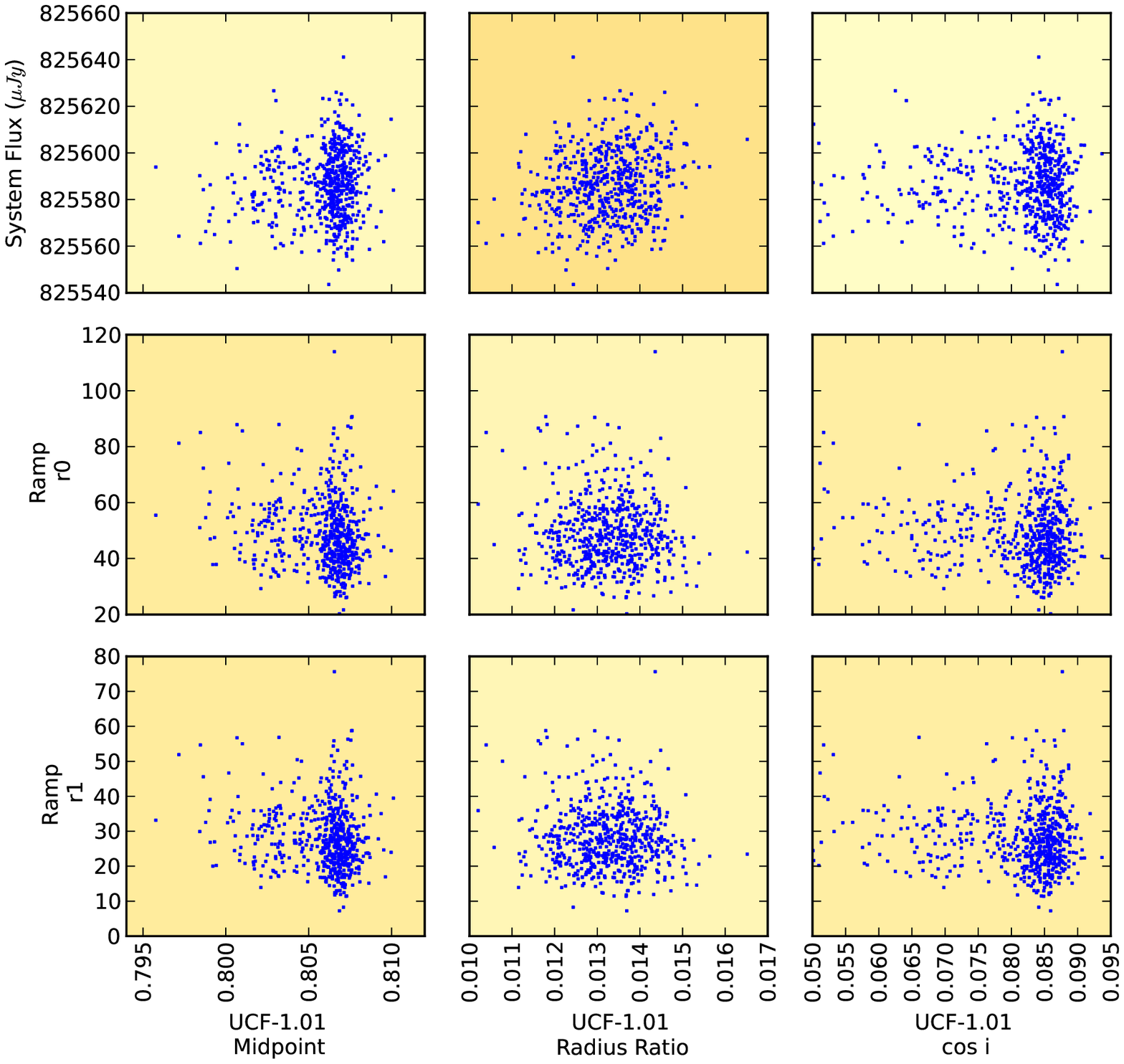}\hfill
\includegraphics[clip,width=0.5\textwidth]{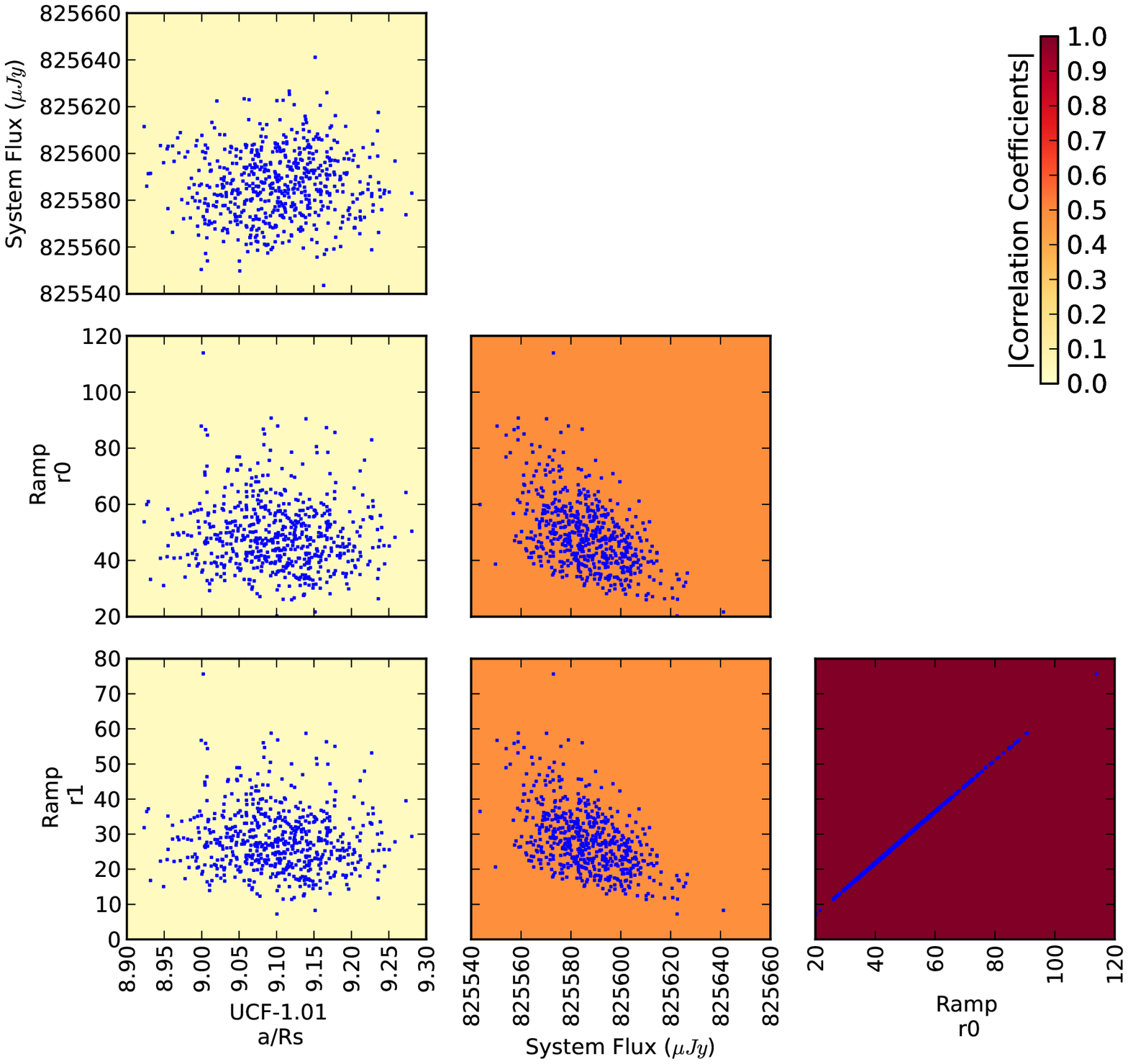}\hfill
\caption{\label{fig:cp22corr}{\small
Correlation plots and histograms for the 2011 July 30 {\em Spitzer} observation containing a transit of UCF-1.01.  We plot every 4000\sp{th} step in the MCMC chain to decorrelate parameter values.
}}
\end{figure}


\clearpage

\section{Best-Fit Parameters}
\label{sec:params}

\begin{table*}[ht]
\caption{\label{tab:params} 
Best joint-fit light-curve parameters}
{\scriptsize
\atabon\strut\hfill
\begin{tabular}{rr@{\,{\pm}\,}lr@{\,{\pm}\,}lr@{\,{\pm}\,}lr@{\,{\pm}\,}lr@{\,{\pm}\,}lr@{\,{\pm}\,}l}
    \hline
    \hline
    Parameter                                                        &   \mctc{ 2010 January 28  }   &   \mctc{   2010 June 29   }   &   \mctc{ 2011 January 24  }   &   \mctc{ 2011 July 30  }   &   \mctc{   2008 July 14   }   \\
    \hline
    Wavelength ({\microns})                                          &   \mctc{       4.5        }   &   \mctc{       4.5        }   &   \mctc{       4.5        }   &   \mctc{       4.5        }   &   \mctc{       8.0        }   \\
    Array Position (\math{\bar{x}}, pix)                             &   \mctc{      14.69       }   &   \mctc{      14.94       }   &   \mctc{      14.63       }   &   \mctc{      14.70       }   &   \mctc{      14.54       }   \\
    Array Position (\math{\bar{y}}, pix)                             &   \mctc{      14.92       }   &   \mctc{      25.31       }   &   \mctc{      15.20       }   &   \mctc{      14.98       }   &   \mctc{      14.52       }   \\
    Position Consistency\tablenotemark{a} (\math{\delta\sb{x}}, pix) &   \mctc{      0.0015      }   &   \mctc{      0.0025      }   &   \mctc{      0.0045      }   &   \mctc{      0.0041      }   &   \mctc{      0.0055      }   \\
    Position Consistency\tablenotemark{a} (\math{\delta\sb{y}}, pix) &   \mctc{      0.0011      }   &   \mctc{      0.0039      }   &   \mctc{      0.0028      }   &   \mctc{      0.0024      }   &   \mctc{      0.0052      }   \\
    Aperture Size (pix)                                              &   \mctc{       2.25       }   &   \mctc{       5.00       }   &   \mctc{       5.25       }   &   \mctc{       5.00       }   &   \mctc{       3.75       }   \\
    Inner Sky Annulus (pix)                                          &   \mctc{       7.0        }   &   \mctc{       10.0       }   &   \mctc{       10.0       }   &   \mctc{       10.0       }   &   \mctc{       7.0        }   \\
    Outer Sky Annulus (pix)                                          &   \mctc{       15.0       }   &   \mctc{       30.0       }   &   \mctc{       30.0       }   &   \mctc{       30.0       }   &   \mctc{       15.0       }   \\
    System Flux, \math{F\sb{s}} (\micro Jy)                          &         841090 & 100          &        871540 & 30            &        819510 & 30            &          825590 & 15          &        315195 & 7             \\
    GJ 436b Tr. Midpt.\tablenotemark{b} (MJD$_{\rm TDB}$)          &   \mctc{        --        }   &   \mctc{        --        }   &   \mctc{        --        }   &   \mctc{        --        }   &    4661.50365 & 0.00012       \\
    GJ 436b \math{R\sb{p}/R\sb{\star}}                               &   \mctc{        --        }   &   \mctc{        --        }   &   \mctc{        --        }   &   \mctc{        --        }   &        0.0830 & 0.0006        \\
    GJ 436b $\cos i$                                                 &   \mctc{        --        }   &   \mctc{        --        }   &   \mctc{        --        }   &   \mctc{        --        }   &         0.066 & 0.002         \\
    GJ 436b \math{a/R\sb{\star}}                                     &   \mctc{        --        }   &   \mctc{        --        }   &   \mctc{        --        }   &   \mctc{        --        }   &          13.0 & 0.3           \\
    GJ 436b Ecl. Midpt.\tablenotemark{b} (MJD$_{\rm TDB}$)         &   \mctc{        --        }   &   \mctc{        --        }   &   \mctc{        --        }   &   \mctc{        --        }   &      4660.417 & 0.003         \\
    GJ 436b Ecl. Midpt.\tablenotemark{b} (MJD$_{\rm TDB}$)         &   \mctc{        --        }   &   \mctc{        --        }   &   \mctc{     5585.7747    }   &   \mctc{        --        }   &      4663.053 & 0.003         \\
    GJ 436b Ecl. Duration (\math{t\sb{\rm 4-1}}, hrs)  &   \mctc{        --        }   &   \mctc{        --        }   &   \mctc{        1.00      }   &   \mctc{        --        }   &          1.02 & 0.13          \\
    GJ 436b Eclipse Depth (ppm)                                      &   \mctc{        --        }   &   \mctc{        --        }   &            18 & 28            &   \mctc{        --        }   &           500 & 60            \\
    GJ 436b $T_{\rm b}$ (K)                                          &   \mctc{        --        }   &   \mctc{        --        }   &           540 & 80            &   \mctc{        --        }   &           700 & 30            \\
    GJ 436b Amplitude, $s\sb{0}$ (ppm)                               &   \mctc{        --        }   &   \mctc{        --        }   &   \mctc{        --        }   &   \mctc{        --        }   &           100 & 40            \\
    GJ 436b Offset\tablenotemark{b}, $s\sb{1}$ (MJD$_{\rm TDB}$)     &   \mctc{        --        }   &   \mctc{        --        }   &   \mctc{        --        }   &   \mctc{        --        }   &       4660.39 & 0.19          \\
    UCF-1.01 Midpt.\tablenotemark{b} (MJD$_{\rm TDB}$)                & \mctc{5225.090$^{+0.004}_{-0.005}$} & \mctc{5376.7078$^{+0.0014}_{-0.0021}$} & \mctc{5585.6889$^{+0.0020}_{-0.0018}$} & \mctc{5772.8069$^{+0.0009}_{-0.0029}$} &      4662.328 & 0.013         \\
    UCF-1.01 \math{R\sb{p}/R\sb{\star}}                               &        0.0138 & 0.0009              &        0.0138 & 0.0009                 &        0.0138 & 0.0009                 &        0.0138 & 0.0009                 &         0.010 & 0.003         \\
    UCF-1.01 $\cos i$                                                 & \mctc{0.084$^{+0.003}_{-0.013}$}    & \mctc{0.084$^{+0.003}_{-0.013}$}       & \mctc{0.084$^{+0.003}_{-0.013}$}       &   \mctc{0.084$^{+0.003}_{-0.013}$}     &         0.084 & 0.003         \\
    UCF-1.01 \math{a/R\sb{\star}}                                     &          9.10 & 0.07                &          9.10 & 0.07                   &          9.10 & 0.07                   &          9.10 & 0.07                   &          9.10 & 0.06          \\
    UCF-1.02 Midpt.\tablenotemark{b} (MJD$_{\rm TDB}$)               &      5225.026 & 0.003               & \mctc{5376.568$^{+0.003}_{-0.007}$}    &   \mctc{        --        }            &   \mctc{        --        }            &   \mctc{        --        }   \\
    UCF-1.02 Transit Depth (ppm)                                     &           186 & 30                  &           186 & 30                     &   \mctc{        --        }            &   \mctc{        --        }            &   \mctc{        --        }   \\
    UCF-1.02 Duration (\math{t\sb{\rm 4-1}}, hrs)                    &    \mctc{1.05$^{+0.22}_{-0.11}$}    &    \mctc{1.04$^{+0.22}_{-0.11}$}       &   \mctc{        --        }            &   \mctc{        --        }            &   \mctc{        --        }   \\
    UCF-1.02 Ingress (\math{t\sb{\rm 2-1}}, hrs)                     &          0.06 & 0.03                &          0.06 & 0.03                   &   \mctc{        --        }            &   \mctc{        --        }            &   \mctc{        --        }   \\
    UCF-1.02 Egress (\math{t\sb{\rm 4-3}}, hrs)                      &          0.06 & 0.03                &          0.06 & 0.03                   &   \mctc{        --        }            &   \mctc{        --        }            &   \mctc{        --        }   \\
    Ramp, \math{r\sb{0}}                                             &           7.0 & 2.0                 &   \mctc{      0           }            &            22 & 12                     &            44 & 11                     &          18.1 & 1.6           \\
    Ramp, \math{r\sb{1}}                                             &          -8.4 & 0.6                 &   \mctc{      0           }            &             6 & 8                      &            25 & 8                      &          -0.6 & 0.5           \\
    Ramp, \math{r\sb{2}}                                             &       -0.0008 & 0.0003              &        0.0020 & 0.0003                 &   \mctc{      0           }            &   \mctc{      0           }            &      -0.00012 & 0.00006       \\
    Ramp, \math{r\sb{3}}                                             &   \mctc{        0         }         &   \mctc{       0.5        }            &   \mctc{        0         }            &   \mctc{        0         }            &   \mctc{       1.5        }   \\
    TIDe                                                             &   \mctc{       Yes        }   &   \mctc{       No         }   &   \mctc{       No         }   &   \mctc{       No         }   &   \mctc{       No         }  \\
    BLISS Map [\math{M(x,y)}]                                        &   \mctc{       Yes        }   &   \mctc{       Yes        }   &   \mctc{       Yes        }   &   \mctc{       Yes        }   &   \mctc{       Yes        }  \\
    Effective Sample Size (ESS)                                      &   \mctc{       341        }   &   \mctc{       702        }   &   \mctc{       398        }   &   \mctc{       675        }   &   \mctc{       415        }   \\
    Minimum \# of Points Per Bin                                     &   \mctc{        6         }   &   \mctc{        4         }   &   \mctc{        8         }   &   \mctc{        6         }   &   \mctc{        4         }   \\
    Total Frames                                                     &   \mctc{      488960      }   &   \mctc{      49536       }   &   \mctc{      51712       }   &   \mctc{      36160       }   &   \mctc{      588480      }   \\
    Rejected Frames (\%)                                             &   \mctc{     0.178        }   &   \mctc{     0.527        }   &   \mctc{     0.673        }   &   \mctc{     0.465        }   &   \mctc{     0.393        }   \\
    Frames Used\tablenotemark{c}                                     &   \mctc{      477106      }   &   \mctc{      48777       }   &   \mctc{      44728       }   &   \mctc{      35172       }   &   \mctc{      583049      }   \\
    Free Parameters                                                  &   \mctc{        6         }   &   \mctc{        10        }   &   \mctc{        5         }   &   \mctc{        4         }   &   \mctc{        18        }   \\
    AIC Value                                                        &   \mctc{     605808       }   &   \mctc{     605808       }   &   \mctc{     605807       }   &   \mctc{     605808       }   &   \mctc{     583067       }   \\
    BIC Value                                                        &   \mctc{     606091       }   &   \mctc{     606091       }   &   \mctc{     606090       }   &   \mctc{     606091       }   &   \mctc{     583270       }   \\
    SDNR                                                             &   \mctc{    0.00535600    }   &   \mctc{    0.00253643    }   &   \mctc{    0.00257029    }   &   \mctc{    0.00258144    }   &   \mctc{    0.00508140    }   \\
    Uncertainty Scaling Factor                                       &   \mctc{     0.31734      }   &   \mctc{    0.17676       }   &   \mctc{     1.06515      }   &   \mctc{     0.98102      }   &   \mctc{     1.04102      }   \\
    Photon-Limited S/N (\%)                            &   \mctc{       84.3       }   &   \mctc{       82.2       }   &   \mctc{       84.0       }   &   \mctc{       83.2       }   &   \mctc{       84.3       }   \\
    \hline
\end{tabular}}
\hfill\strut\ataboff
\tablenotetext{1}{RMS frame-to-frame position difference.}
\tablenotetext{2}{MJD = BJD - 2,450,000.}
\tablenotetext{3}{We exclude frames during instrument/telescope settling, for insufficient points at a given knot, and for bad pixels in the photometry aperture.}
\end{table*}

\end{appendices}

\end{document}